\documentclass[10pt,prb,aps,twocolumn,showpacs,floatfix,superscriptaddress]{revtex4}
\usepackage{graphicx}
\usepackage{amssymb}
\usepackage{color}
\usepackage{amsmath} 
\newcommand{\be}{\begin{equation}}
\newcommand{\ee}{\end{equation}}

\begin{document}

\title{Velocity of excitations in ordered, disordered and critical antiferromagnets}

\author{Arnab Sen} 
\affiliation{Department of Theoretical Physics, Indian Association for the Cultivation of Science, Jadavpur, Kolkata 700032, India}

\author{Hidemaro Suwa}
\affiliation{Department of Physics, University of Tokyo, Tokyo, Japan}
\affiliation{Department of Physics, Boston University, 590 Commonwealth Avenue, Boston, Massachusetts 02215, USA}

\author{Anders W. Sandvik}
\affiliation{Department of Physics, Boston University, 590 Commonwealth Avenue, Boston, Massachusetts 02215, USA}
\affiliation{Department of Physics, National Taiwan University, Taipei 10607, Taiwan} 

\begin{abstract}
We test three different approaches, based on quantum Monte Carlo simulations, for computing the velocity
$c$ of triplet excitations in antiferromagnets. We consider the standard $S=1/2$ one- and two-dimensional
Heisenberg models, as well as a bilayer Heisenberg model at its critical point. Computing correlation
functions in imaginary time and using their long-time behavior, we extract the lowest excitation energy
versus momentum using improved fitting procedures and a generalized moment method. The velocity is then
obtained from the dispersion relation. We also exploit winding numbers to define a cubic space-time geometry,
where the velocity is obtained as the ratio of the spatial and temporal lengths of the system when all winding
number fluctuations are equal. The two methods give consistent results for both ordered and critical systems,
but the winding-number estimator is more precise. For the Heisenberg chain, we accurately reproduce the exactly
known velocity. For the two-dimensional Heisenberg model, our results are consistent with other recent
calculations, but with an improved statistical precision; $c=1.65847(4)$. We also use the hydrodynamic relation
$c^2=\rho_s/\chi_\perp(q\to 0)$ between $c$, the spin stiffness $\rho_s$, and the transversal susceptibility
$\chi_\perp$, using the smallest non-zero momentum $q=2\pi/L$. This method also is well controlled in two
dimensions, but the cubic criterion for winding numbers delivers better numerical precision. In one dimension
the hydrodynamic relation is affected by logarithmic corrections which make accurate extrapolations difficult.
As an application of the winding-number method, for the quantum-critical bilayer model our high-precision
determination of the velocity enables us to quantitatively test, at an unprecedented level, field-theoretic
predictions for low-temperature scaling forms where $c$ enters. We find agreement to within $3\%$ with $1/N$ expansions for
the coefficients of the leading susceptibility and specific heat forms; $\chi \sim aT$ and $C \sim bT^2$.
\end{abstract}

\date{\today}


\maketitle

\section{Introduction}

Ordered quantum antiferromagnets exhibit linear dispersions of their elementary spin-wave (magnon) excitations and  the associated velocity 
$c$ is an important parameter characterizing such systems. The prototypical two-dimensional Heisenberg model has an ordered ground state 
and its low-energy magnon spectrum is well described by spin wave theory.\cite{manousakis91} When $1/S$ corrections are properly taken into account, 
the velocity and other properties computed within this approximation for the most extreme (and interesting) case of the spin $S=1/2$ model agrees to 
within $\approx 1\%$ with results of quantum Monte Carlo (QMC) calculations, which can be considered exact to within statistical errors if sufficiently 
large lattices are used for extrapolations to the thermodynamic limit.\cite{runge92,wiese94,sandvik97,sandvik10a,jiang11a} This good agreement 
can be traced to the fact that the ground state is strongly ordered, the sublattice magnetization being reduced by quantum fluctuations by only about 
$40\%$ from the classical value. Upon introducing other interactions which enhance the quantum fluctuations and suppress the order, the quantitative 
predictive power of spin-wave calculations rapidly deteriorates and the quantum fluctuations have to be treated in more sophisticated 
ways. \cite{singh88,millis93,chubukov95,sandvik94,sandvik99a} An extreme case is when a system is driven to criticality. The low-energy critical 
excitations are still linearly dispersing but the corresponding velocity cannot be reliably calculated in any simple theoretical manner. 
Lastly, quantum-disordered antiferromagnets also have propagating triplet excitations, which are gapped and often called triplons. 

In this paper we will discuss three very different ways to extract the velocity of the elementary excitations of quantum spin models based on ground-state 
projector and finite-temperature QMC simulations. We consider one-dimensional (1D) and two-dimensional (2D) ordered, disordered and critical quantum 
antiferromagnets. Using imaginary-time dependent spin correlation 
functions, the long-time behavior computed at different momenta contain information on the dispersion relation, from which the velocity can be extracted 
if the limits of the system size going to infinity and the momentum going to zero are treated correctly. One can also in some cases extract the velocity 
in a simpler, indirect way using winding number fluctuations in the space-time representation of the system sampled in the QMC calculations.\cite{kaul08,jiang11b} 
We will develop stable procedure based on these two approaches and compare the results for a number of different cases. In addition, we also test the well
known hydrodynamic relationship $c^2=\rho_s/\chi_\perp$ for an antiferromagnetic state,\cite{halperin69} where $\chi_\perp$ is the transversal magnetic susceptibility 
and $\rho_s$ the spin stiffness. 

We first successfully test the methods on both 1D and 2D systems for which the velocity (of spinons and spin waves, respectively) is previously 
known, and thereafter study a bilayer system, both in its paramagnetic phase and at its quantum-critical point. In the latter case we subsequently use our 
high-precision estimate of $c$ to investigate in detail, to our knowledge at an unprecedented level of control, the reliability of finite-temperature 
quantum-critical scaling forms for the magnetic susceptibility and the specific heat.\cite{chakravarty89,chubukov94} 

Before turning to the calculations and results, in Sec.~\ref{sec:effective} we first provide some more background on the utility of carrying out 
precise determinations of the velocity in quantum spin systems, focusing on quantum-criticality in dimerized antiferromagnets. In \ref{sec:objectives} 
we provide a brief summary of the different calculations to be presented and in \ref{sec:outline} we outline the organization of the rest of the paper.

\subsection{Effective low-energy descriptions}
\label{sec:effective}

Quantum field theories are often used to describe universal low-energy properties 
of quantum magnets.\cite{chakravarty89,chubukov94} Numerical techniques, such as QMC simulations, can be used to extract system-dependent 
parameters appearing in various predicted forms of physical quantities at low energy. Such an approach of combining quantum field theory and numerics 
has been established over the past several years for certain types of quantum phase transitions in 2D systems.\cite{kaul13, kim98} Most well studied are 
transitions in {\it dimerized} models, where for weak inter-dimer coupling there is a tendency to singlet formation on the dimers, leading to a 
quantum phase transition into a quantum paramagnet at a critical ratio of the inter- and intra-dimer 
couplings.\cite{singh88,sandvik94,shevchenko00,matsumoto01,brenig06,wang06,wenzel09,fritz11} A similar transition (of the same universality class) also take 
place if the dimers are replaced by some other unit cell of an even number of spins on which the single-cell ground state is a 
singlet.\cite{troyer96,albuquerque08} The many studies of these systems have shown rather convincingly that the phase transition is in the 
3D O$(3)$ universality class, in agreement with field-theory descriptions based on the non-linear sigma model.\cite{chubukov94,chakravarty89} 
Other properties associated with quantum-criticality are also well captured by the field theory,\cite{chubukov94} e.g., the uniform magnetic 
susceptibility $\chi$ is linear in temperature at the critical coupling ratio, $\chi = aT/c^2$, and the specific heat grows 
quadratically; $C = bT^2/c^2$ at low $T$. Perhaps surprisingly, however, the values of the prefactors $a$ and $b$ have still not been 
tested quantitatively in a completely unbiased manner.\cite{sandvik11} This may be largely because, as indicated above, they depend on the 
velocity $c$ of the critically damped spin waves (and on no other low-energy parameter), but this parameter has not yet been independently calculated 
to high precision for model systems (by first-principles methods not depending on other field-theory predictions). As a demonstration of the value of
determining $c$ to high precision, in this paper we will also provide a test case of a detailed comparison with field theory predictions
for one of the prototypical dimerized models; the Heisenberg bilayer.

\subsection{Technical and Physics Objectives}
\label{sec:objectives}

The first aim of the present paper is to systematically test three completely different ways of extracting the velocity of elementary excitations 
based on QMC calculations in the following ways: (i) Using the momentum dependent gaps extracted from imaginary-time dependent spin correlation functions. 
(ii) By monitoring spatial and temporal winding number fluctuations, which are proportional to the spin stiffness and the uniform susceptibility, respectively, and 
adjusting the space-time aspect ratio $L/\beta$ such that these fluctuations are equal. At this special inverse temperature $\beta^*(L)$ the ratio of the spatial 
and temporal lengths $L/\beta^*$ should equal $c$.\cite{kaul08,jiang11b} (iii) Exploiting the hydrodynamic relation $c^2=\rho_s/\chi_\perp$, where one has pay attention to the 
fact that $\chi_\perp \to 0$ when the temperature $T \to 0$ in a finite system, while the spin stiffness remains nonzero. We discuss an approach to 
circumvent this problem based on $\chi_\perp(q)$ where the momentum $q$ is small but non-zero.

The method (i) is in principle very direct, being connected to the fundamental definition of $c$ in the dispersion relation of the lowest-energy excitation versus 
momentum. However, the precise determination of the momentum-dependent gap is in practice complicated by the presence of a continuum of 
excitations above the lowest energy. In some cases, as we will show, one also has to take great care with the way the thermodynamic limit
and zero-momentum limits are taken. The method (ii) is rather simple and the only uncertainty is introduced by a final extrapolation of the finite-size
velocity definition $L/\beta^*$ to the thermodynamic limit. However, as far as we are aware, the correctness of this approach has not been formally proven, except 
for the case of a long-range ordered antiferromagnet, and the method has not been widely used.\cite{kaul08,jiang11b} We here confirm that 
the  method continues to give the correct velocity also when the system is critical, even in the case of the $S=1/2$ Heisenberg chain, where the 
low-energy excitations are not even spin-waves but deconfined spin-$1/2$ carrying topological defects (``spinons''). The method (iii) based
on generalized hydrodynamics is simple to apply and we will argue that it as well continues to work also in critical systems.

Comparing methods (i)-(iii) for the Heisenberg chain as well as for the well-studied 2D Heisenberg model gives us insights into how to best apply the methods 
in practice. After these tests, we study the bilayer Heisenberg model at its quantum-critical point and in its paramagnetic regime, using methods (i) and (ii) Also in this 
case we find good agreement between the two methods at the critical point, but one has to be more careful when defining the velocity based on gaps for finite systems, 
because of a slower convergence of the gaps to their infinite-size values than in the ordered state. We also explicitely show that the winding number estimator 
gives an {\it incorrect} velocity in the gapped paramagnetic phase.

The second aim of the paper connects to the low-energy filed-theory description discussed in Sec.~\ref{sec:effective}---to compute a high-precision value for $c$ for the 
quantum-critical Heisenberg bilayer model, and to use this value to reliably test the field-theory predictions for $\chi(T)$ and $C(T)$. While many such tests have been 
performed in the past, on the bilayer\cite{sandvik94,shevchenko00,brenig06} as well as other \cite{troyer96} quantum-critical 2D spin systems, the past studies were not able 
to completely quantitatively test the level of agreement with the existing large-$N$ field theory predictions, because an independent, unbiased determination of 
$c$ was lacking. This obstacle is overcome by the reliable calculation of $c$ in this paper.

\subsection{Outline of the Paper}
\label{sec:outline}

The rest of the paper is organized as follows: In Sec.~\ref{sec:gaps} we discuss calculations of imaginary-time correlation functions within a 
ground-state projector QMC method formulated in the basis of valence bonds. We discuss our methods to extract the lowest momentum-dependent gaps from 
the long-time behavior of these correlations. In addition to fitting a leading exponential decay with additional corrections to account for higher
excitations, we also present a generalization to ground-state projection QMC data of a systematic high-order moment approach recently introduced for 
use with finite-temperature QMC simulations.\cite{SuwaT2014} We compare our results with exact solutions for small Heisenberg chains as well as with 
the rigorously known velocity of this model in the thermodynamic limit. In Sec.~\ref{sec:winding} we discuss the determination of $c$ based on winding 
numbers and demonstrate the method using the Heisenberg chain. These finite-temperature calculations are carried out with the stochastic series 
expansion (SSE) QMC method. In Sec.~\ref{sec:hydromethod} we discuss details of the hydrodynamic relationship generalized to finite system size
and test it using SSE calculations for the Heisenberg chain. In Sec.~\ref{sec:2D} we present further tests of all three methods for the standard 2D 
Heisenberg model. We also show that the scaling of the triplet gap at momentum $\mathbf{k} = (\pi,\pi)$ is entirely consistent with quantum rotor-states 
carrying spin $S=1$. In Sec.~\ref{sec:bilayer} we discuss results for the critical and disordered bilayer systems, including the comparison of $c$ 
determined using both methods. The analysis of the quantum-critical susceptibility and specific heat computed using large-scale SSE calculations is presented in Sec.~\ref{sec:bilayerT>0}.
We briefly summarize and discuss our results further in Sec.~\ref{sec:discussion}.

\section{Momentum-dependent spin gaps}
\label{sec:gaps}

Here we discuss how to use zero temperature ($T=0$) projector QMC methods to calculate imaginary-time correlation functions which are then used to 
extract the triplet gap (with respect to the ground state energy) as a function of momentum $\mathbf{k}$. Applying the imaginary-time evolution
operator ${\rm e}^{-\beta H}$ to a trial state leads to the ground state when $\beta \to \infty$. Alternatively, one can use a high power $H^m$ of
the Hamiltonian and reach the ground state when $m \to \infty$. In practice QMC simulation methods based on these two operators are very similar, but 
the exponential form allows more direct access to the standard imaginary time. As we will show, this connection is not needed if only the excitation energies 
are of interest, and then one can use the slightly faster power method. We will use both approaches here, with two different ways of analyzing the
correlation functions.

\subsection{$H$-power projection}

This QMC approach is based on projection with a sufficiently high power of the hamiltonian $H$ on a trial state $|\psi_t \rangle$. The process can be 
conveniently expressed in the energy eigenbasis of $H$, leading to the following expression for the dependence on $m$: 
\begin{eqnarray}
(-H)^m |\psi_t \rangle = c_0 (-E_0)^m \left[ |0 \rangle +
  \sum_{n=1}^{\Lambda} \frac{c_n}{c_0} \left(\frac{E_n}{E_0} \right)^m |n \rangle \right]\hskip-1mm.~~~
\end{eqnarray}
Here $|n \rangle$, $n=0,\cdots,\Lambda-1$ are the energy eigenstates of $H$ and $E_0$ is assumed to be negative, with its absolute 
value $|E_0|$ being the largest in magnitude of all the energies (which can always be achieved by adding a suitable constant to the hamiltonian). 
Then, if the expansion coefficient $c_0 \neq 0$ (which in practice is essentially guaranteed for any reasonable choisce of $\psi_t\rangle$), we 
have $(-H)^m |\psi_t \rangle \propto |0\rangle$ for large $m$, and the expectation value of an operator $O$ at $T=0$ can be written as
\begin{equation}
\langle O \rangle  = \frac{\langle \psi_t| (-H)^m O (-H)^m |\psi_t \rangle}{\langle \psi_t| (-H)^{2m} |\psi_t \rangle}.
\end{equation}
For the SU(2) invariant spin models considered in this paper, this form of the expectation value can be evaluated by importance-sampling, using a 
formulation of the projector QMC method in the non-orthogonal valence bond basis.\cite{sandvik05} An efficient way of sampling the contributions to 
$(-H)^{2m}$, very similar to ``operator-loop'' updates developed within the finite-$T$ SSE method,\cite{sandvik99} can be formulated using loop updates 
in a combined space of spin components ($S^z$) and valence bonds.\cite{sandvikevertz10} The trial state is also expressed using valence bonds, in the 
form of an amplitude-product state.\cite{liang88} The details of the state (the form of the amplitudes) are not important, as the good convergence to 
the ground state is observed even if the state is not optimized.\cite{sandvikevertz10} 

We will use this valence-bond variant of the projector QMC method for computing appropriate imaginary-time dependent correlation functions.   
For technical details on the sampling methods we refer to Ref.~\onlinecite{sandvik10a}. Below we will focus on the definition of the correlation
functions we study and how we process them to extract the velocity.

\subsubsection{Imaginary-time correlations}

We consider correlation functions of the following form: 
\begin{equation}
C_A(t) = \frac{\langle 0 | A^{\dagger} (-H)^t A |0 \rangle }{\langle 0| (-H)^t
|0 \rangle } = \sum_{n=0}^{\Lambda-1} \left |\frac{E_n}{E_0} \right |^t
|\langle n|A|0 \rangle |^2,
\label{catdef}
\end{equation}
where $t$ is an integer which can be related to imaginary time \cite{sandvik92} and we will loosely refer to it by this term. More precisely, 
$t/N$, where $N$ is the system size, is proportional to imaginary time $\tau$ in the sense of the standard Schr\"odinger time evolution 
operator ${\rm e}^{-\tau H}$, as we will explicitly show below. From $C_A(t)$, we further define
\begin{equation}
Q_A(t) = \frac{C_A(t)-|\langle 0|A|0 \rangle|^2}{C_A(0)-|\langle 0|A|0 \rangle|^2},
\label{QAdef}
\end{equation}
and note that $Q_A(0) = 1$ and $Q_A( t \rightarrow \infty) \rightarrow 0$. For large $t$, 
we have 
\begin{equation}
Q_A(t) \rightarrow \left(\frac{|\langle 1| A |0 \rangle|^2}{\langle 0|
    A^{\dagger}A|0 \rangle -|\langle 0| A |0 \rangle|^2 } \right) \left(1-
  \frac{\Delta}{|E_0|} \right)^t,
\label{longtimeQ}
\end{equation}
where $\Delta = E_1-E_0$ is the energy gap between the first excited state (connected to the ground state by the operator $A$) and the
ground state. To directly show the relationship between $t$ and imaginary time, we can introduce $\tau=t/|E_0|$, then write $E_0=Ne_0$, 
and in the limit of large $N$ have
\begin{equation}
Q_A(\tau=t/|E_0|)  \propto \left(1-\frac{\Delta}{N|e_0|} \right)^{N|e_0|\tau} \to ~{\rm e}^{-\Delta\tau},
\label{longtimeQ2}
\end{equation}
which is the familiar form of the asymptotic decay of an imaginary-time dependent correlation function. Thus, we have shown that, indeed, 
$\tau \propto t/N$. We will not explicitly need to make use of the relationship between $t$ and $\tau$ here, however, and we will 
continue to use $t$ as the ``time'' parameter with the $H$-power method.

We can appropriately choose the operator $A$ so that it excites the ground state $|0 \rangle$ into a state with desired quantum numbers.  
The ground state of the unfrustrated hamiltonians considered here are total spin singlets with momentum $\mathbf{k} = 0$ on a finite lattice 
with an even number of spins and periodic boundary conditions (in one dimension only when the number $N$ of sites is a multiple of four---for
other even $N$ the momentum is $\pi$).\cite{sandvik10b} Since we are interested in triplet excitations, we can use the
simple Fourier-transformed spin operator,
\begin{equation}
 A(\mathbf{k})
= S^z(\mathbf{k}) = \frac{1}{\sqrt{N}}\sum_\mathbf{r} {\rm e}^{i \mathbf{k} \cdot \mathbf{r}} S^z(\mathbf{r}),
\label{akszdef}
\end{equation}
to create an $S=1$ state with momentum $\mathbf{k}$ and $S^z=0$ when acting on the ground state.
Thus, the following imaginary-time correlation function
\begin{eqnarray}
C_{\mathbf{k}}(t) &=& \frac{\langle 0 | A(-\mathbf{k}) (-H)^t A(\mathbf{k})
  |0\rangle }{\langle 0| (-H)^t |0 \rangle } \\
                &=& \frac{ \langle \psi_t| (-H)^{2m - p -t} A (-\mathbf{k}) (-H)^t A (-H)^p
  |\psi_t \rangle }{\langle \psi_t | (-H)^{2m} |\psi_t \rangle } \nonumber 
\end{eqnarray}
allows us to directly measure the triplet excitation gap $\Delta(\mathbf{k})$
as a function of the momentum. The second line in the above equation
explicitly shows the form used in the projector QMC calculations, where both
$2m-p-t$ and $p$ are assumed to be large enough to achieve projection to the ground
state for the system sizes considered. 

In practice, we will use projection powers $m = 16 p_r N$ and, to achieve good ground-state convergence, require that $2m-p-t$ and $p$ both are larger than 
$15 p_r N$, where typically $p_r =8$ ($16$ or higher in some cases). These choices are motivated by extensive tests indicating that no detectable systematical 
errors remain. The values of $t$ are restricted to be multiples of $N/4$ in our simulations. Since the correlation functions at the different $t$ values are 
measured in the same simulation, the $C_{\mathbf{k}}(t)$ data are correlated and measuring at shorter $t$ intervals does not significantly
increase the amount of statistical information in the data set.

\subsubsection{Extracting the gap}

We here use two different ways to extract the lowest triplet gap from the correlation function $C_{\mathbf{k}}(t)$. 
Since $\langle 0|A(\mathbf{k})|0 \rangle=0$ for $\mathbf{k} \not=0$, Eq.~(\ref{QAdef}) reduces to
\begin{equation}
Q_{\mathbf{k}}(t) =\frac{C_{\mathbf{k}}(t)}{C_{\mathbf{k}}(0)}.
\label{Qtdef2}
\end{equation}
In principle, the gap can be extracted by monitoring the long-time behavior, given by the form (\ref{longtimeQ}). A systematic way to extract the
gap without performing any curve fits is to consider the ratio of $Q_{\mathbf{k}}(t)$ at two different times separated by some interval; e.g., by $N/4$ 
operations:
\begin{equation}
R_{\mathbf{k}}(t) = |E_0| \left(1-
  \left[\frac{Q_{\mathbf{k}}(t+N/4)}{Q_{\mathbf{k}}(t)}
  \right]^{4/N} \right). 
\label{rdef}
\end{equation}
Note that $R_{\mathbf{k}}(t)
\rightarrow \Delta(\mathbf{k})$ when $t$ is large enough, which follows from the long-time behavior of $Q_{\mathbf{k}}(t)$ in Eq.~(\ref{longtimeQ}). However, 
in our QMC calculations, it is not always possible to reach perfect convergence of this gap estimate for all $\mathbf{k}$, because the relative statistical 
errors often become too large already for moderately large $t$. This problem is related to the existence of a continuum (for large $N$) of states above the
gap, due to which the pure exponential decay cannot be easily observed in practice. 

We thus use another method to estimate the value of the gap based on the entire available set of correlation functions. This scheme is more reliable 
than the ratio scheme when data (with small relative errors) are not available for large values of imaginary time. It is clear from Eq.~(\ref{catdef})
that one can define a positive-definite spectral function $A_{\mathbf{k}}(\omega)$ to fit the normalized imaginary-time correlation $Q_{\mathbf{k}}(t)$ as
\begin{equation}
 Q_{\mathbf{k}}(t) = \int_0^\infty d\omega A_{\mathbf{k}}(\omega) \left(1-\frac{\omega}{|E_0|} \right)^t.
\label{gaps1}
\end{equation}
This is just the analogue for the $H$-power evolution of the standard form, 
\begin{equation}
 G_{\mathbf{k}}(\tau) = \int_0^\infty d\omega S_{\mathbf{k}}(\omega) {\rm e}^{-\omega\tau},
\label{gtsqomega}
\end{equation}
relating the imaginary-time Schr\"odinger evolved correlation function
\begin{equation}
G_{\mathbf{k}}(\tau) = \langle 0| S^z_{-\mathbf{k}} (\tau) S^z_{\mathbf{k}}(0)|0\rangle,
\label{gtaudef}
\end{equation}
to the dynamic spin structure factor $S_{\mathbf{k}}(\omega)$. In Eq.~(\ref{gaps1}) $\omega$ cannot exceed $|E_0|$, following from the fact that we 
have ensured that $E_0$ is the eigenvalue with the largest magnitude. In practice, as in the standard dynamic spin structure factor 
in (\ref{gtsqomega}), the actual dominant spectral weight will be concentrated only to within a window of order $J$. 

For any finite system, $A_{\mathbf{k}}$ is a sum of $\delta$-functions and this can be replaced by a continuum starting at $\omega=\Delta(\mathbf{k})$ 
for a large system (or, in some cases, there is an isolated $\delta$-function at the gap, following by a second gap and then a continuum). 
With $G_{\mathbf{k}}(\tau)$ or $Q_{\mathbf{k}}(t)$ computed using QMC calculations, the respective relations (\ref{gaps1}) or 
(\ref{gtsqomega}) can in principle be inverted using numerical analytic continuation. This procedure is very challenging, however, and it is not easy to 
extract the gap precisely with conventional methods such as the Maximum Entropy method,\cite{jarrell96} though one can extract the main dominant spectral 
features (and we note that progress in this regard has been made very recently \cite{sandvik15}). Here our goal is merely to extract the gap, and instead 
of trying to reproduce the full spectral function we model the excitations by just a small number of $\delta$-functions. With the precision of typical QMC 
data, $Q_{\mathbf{k}}(t)$ can be normally fitted very well with just a few $\delta$-functions (typically $3$ to $5$) over the full range of accessible times 
$t$. With this procedure, we expect the location $\omega_1$ of the lowest gap to accurately reproduce $\Delta(\mathbf{k})$, while the 
higher $\delta$-functions represent approximately the contributions of the continuum. In this fitting procedure, the extracted $\omega_1$ 
is to some extent affected by contributions of the higher states but does not change significantly when increasing the number of $\delta$-functions.
We can therefore quite reliably extract the lowest gap, but not higher ones unless they are separated by significant subsequent gaps (which is not
expected in the cases of interest here, except well inside the quantum paramagnetic state of the bilayer model).

Given a set of $n$ $\delta$-functions at energies $\omega_i$ with associated amplitudes $A_i$ normalized so that $\sum_i A_i = 1$, one can compute 
the associated time dependent correlation function in analogy with Eq.~(\ref{gaps1}) as 
\begin{equation}
 Q_{\mathbf{k}}(t) = \sum_{i=1}^n A_i \left(1-\frac{\omega_i}{|E_0|} \right)^t.
\label{gaps2}
\end{equation}
Now denoting the corresponding QMC-computed function by $\tilde Q_{\mathbf{k}}(t)$ and their statistical 
error by $\sigma_t$, the goodness of the fit is quantified in the standard way by $\chi^2$, based on a set of $N_t$ time points $\{t\}$:
\begin{equation}
\chi^2 = \frac{1}{N_t} \sum_{\{t\}} \frac{1}{\sigma_t^2} \bigl ( Q_{\mathbf{k}}(t)-\tilde Q_{\mathbf{k}}(t) \bigr )^2. 
\label{chi2def}
\end{equation}
We here use a uniform grid of time points with separation $N/4$ operations, $t=N/4,N/2,...$, up to a point $t_{\rm max}=N_t(N/4)$ where the relative statistical
error of $\tilde Q_{\mathbf{k}}(t)$ exceeds $5 \%$. The choice of cut-off is not very important as, in any case, the noisy data at very large $t$ will not 
affect the fit from the definition of $\chi^2$.

For the extracted gap to be reliable, the contribution of the lowest $\delta$-function to the fit must be significant at the longest times included. To
monitor this long-time weight, we compute the relative contribution of the lowest $\delta$-function, denoted by $A_1(t)$, at the time $t_{\rm max}$ included in the fit:
\begin{equation}
A_1(t_{\rm max}) = \frac{A_1 \left(1- {\omega_1}/{|E_0|}
  \right)^{t_{\rm max}}}{\sum_{i=1}^n A_i \left(1- {\omega_i}/{|E_0|}
  \right)^{t_{\rm max}}}.
\end{equation}
This quantity should approach $1$ for $t \to \infty$ if the lowest $\delta$-function is at the gap. It is close to $1$ in all the fits reported
here, indicating stable extraction of $\omega_1$.

\begin{figure}
\center{\includegraphics[width=7.2cm, clip]{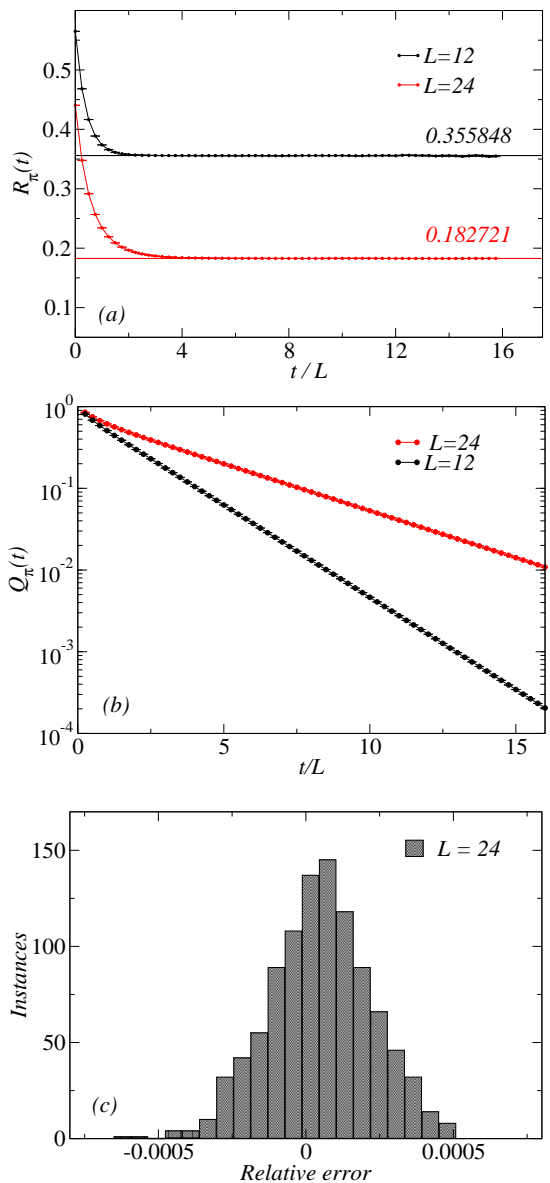}}
\vskip-2mm
\caption{(Color online) (a) Correlation ratio $R_{\pi}(t)$, Eq.~(\ref{rdef}), shown for two Heisenberg chain sizes, $L=12$ and $24$. The lines are the exact 
values of the triplet gap at this wave-vector (with the numerical values indicated as well). (b) The normalized correlation function $Q_{\pi}(t)$ for $L=12$ 
and $L=24$. (c) The distribution of the relative error of the gap $\Delta(\pi)$ extracted by fitting the data for $L=24$ in (b) using three $\delta$-functions.
The histogram was generated from a large number of bootstrap samples of the QMC data. The relative error is calculated as $(\Delta - \Delta_e)/\Delta_e$ 
where $\Delta_e$ is the exact value of the gap. The exact value (corresponding to relative error $0$) is seen to fall within a standard deviation of the 
distribution of the bootstrap samples.}
\label{1dsmallchains}
\vskip-2mm
\end{figure}

The statistical error of the extracted gap is estimated using a bootstrap error analysis. With the underlying QMC data for the
correlation function $C_{\mathbf{k}}(t)$ saved as $M$ bin averages (with typically $M \sim 100-1000$), bootstrap averages are constructed by selecting 
$M$ bins at random (i.e., allowing for the same bin to be selected multiple times). The above fitting procedures are then carried out repeatedly for
a large number of these samples, and the standard deviation of the estimates is the statistical error of the gap in our procedure.  

As already mentioned, the fluctuations of the QMC data at different times are significantly correlated since these are measured in the same 
simulation. A statistically correct treatment of the data would require the inclusion of the full covariance matrix (instead of just its
diagonal elements) in the definition of $\chi^2$. However, much of the covariance is already removed when the time-correlations are normalized 
[by the denominator Eq.~(\ref{Qtdef2})], because the errors are correlated primarily by overall fluctuations in the normalization. Based on test cases, 
including ones reported below, to obtain fully reliable results it is sufficient to use only the diagonal elements of the covariance matrix 
and define $\chi^2$ as in (\ref{chi2def}).  
 
\subsubsection{Tests on the Heisenberg chain}

We here illustrate the gap extraction method described in the previous subsection using
the example of the $S=1/2$ Heisenberg spin chain with periodic boundary conditions, where spins interact with
nearest neighbor exchange constant $J=1$;
\begin{equation}
H = J\sum_{i=1}^L {\bf S}_i \cdot {\bf S}_j.
\end{equation}
First, we compare the results of our numerics for the lowest triplet gap at $k=\pi$ for chain sizes $L=12$ and $24$ with exact diagonalization 
(Lanczos-method) results.  As can be seen in Fig~\ref{1dsmallchains}(a), the quantity $R_{\pi}(t)$ defined in Eq.~(\ref{rdef}) indeed 
converges to the correct gap value in both the cases. Fig~\ref{1dsmallchains}(b) shows the normalized imaginary time correlation function $Q_{\pi}(t)$ 
for the two system sizes, and Fig~\ref{1dsmallchains}(c) shows the distribution of the gap error obtained using a bootstrap method (i.e., the simulation data 
stored as $M$ bin averages are re-sampled by selecting $M$ bins at random, and the fitting procedures are carried out for each such averaged data set) 
for $L=24$, using a fit to three $\delta$-functions. 

This analysis show that the gap obtained from the fit agrees statistically with the exact result, which here
is within a standard deviation of the distribution obtained in the bootstrapping procedure, and the distribution itself closely matches a normal distribution.
Thus, even with only three $\delta$-functions in the spectrum (which is clearly a much smaller number than what is contained in the full spectrum) we detect 
no systematic errors introduced by the fitted functional form, supporting our assertion that the simplified description of the spectrum does not significantly affect the 
location of the lowest $\delta$-function (the gap).  The number $n$ of $\delta$-functions that should be included in a given case depends on the statistical errors of the 
QMC data and the actual form of the spectral continuum. To determine $n$, we monitor $\chi^2$ as a function of increasing $n$ and stop when no improvements 
in the fit are observed.

\begin{figure}
\center{\includegraphics[width=7cm, clip]{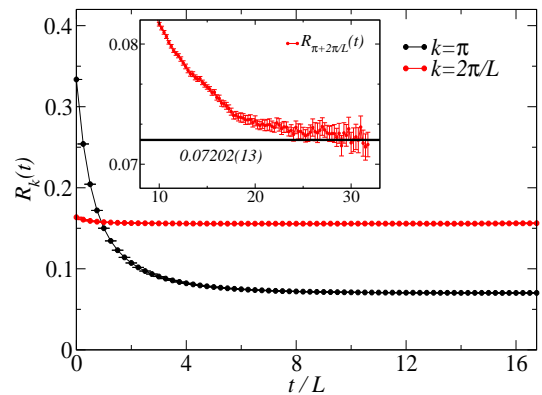}}
\vskip-2mm
\caption{(Color online) Convergence of the correlation ratio $R_k(t)$ to the gap $\Delta(\mathbf{k})$ as a function of imaginary time $t$, shown at  $k=\pi$ and $k=2\pi/L$ 
for a Heisenberg chain of length $L=64$. The inset shows $R_k(t)$ at $k=\pi+2\pi/L$ for a longer chain, $L=200$, where the gap extraction using a fit of $Q_k(t)$ to $5$ 
$\delta$-functions works demonstrably better, giving the result indicated below the horizontal line, with the thickness of the line representing the average gap $\pm$ one error bar.}
\label{L64}
\end{figure}

For the Heisenberg chain, the lower edge of the spectrum of triplet excitations is known rigorously based on the Bethe Ansatz solution.\cite{cloizeaux62}
For momentum $k\to 0$ and $k \to \pi$ the spectrum is linear with velocity $c=\pi/2$. For an infinite chain the triplets are degenerate with 
singlet excitations, due to the fact that the excitations consist of essentially independently propagating deconfined spinons, each carrying spin $1/2$. 
The $2$-spinon continuum is maximally broad at $k=\pi$ and shrinks to zero at $k=0$. This leads to larger contributions to the time correlations  from the 
continuum close to $k=\pi$, which is directly visible in the QMC data for long chains as, e.g., a slower rate of convergence of $R_k(t)$ to a constant. For 
example, $R_{2\pi/L}(t)$ converges much faster to its asymptotic constant value compared to $R_{\pi}(t)$ for a chain of size $L=64$, as shown in Fig.~\ref{L64}. 
In the inset of the same figure we also show results for $k=\pi+2\pi/L$, which is the momentum we use to extract $c$ as discussed further below. Here the chain is
longer, $L=200$, and $R_{k}(t)$ does not converge sufficiently before the error bars become too large. However, the method of fitting a simplified spectral
function to $Q_{\pi}(t)$ still works very well and delivers a gap consistent with $R_{k}(t)$, but with much better statistical precision. We therefore exclusively
uses the fitting method to obtain the results to be discussed next.

\begin{figure}
\center{\includegraphics[width=7cm, clip]{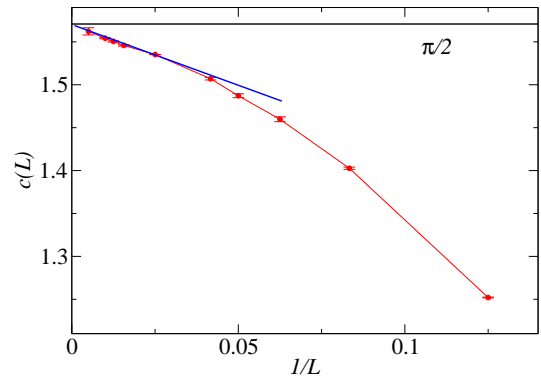}}
\vskip-2mm
\caption{(Color online) Finite-size velocity estimates for the Heisenberg spin chain obtained from triplet gaps at and in the neighborhood
of $k=\pi$ according to Eq.~(\ref{cldef}), shown as a function of the inverse system size. The velocity approaches the known value $c=\pi/2$ (indicated 
by the horizontal line) for large systems. The line through the larger system sizes is a guide to the eye.}
\label{velexact1dpi}
\vskip-2mm
\end{figure}

In a finite chain, the ground state is non-degenerate and has momentum $k=0$. However, there is also a quasi-degenerate state with $k=\pi$, which is
obtained by adding an Umklapp to the true ground state, and this state becomes exactly degenerate with the $k=0$ ground state when $L \rightarrow \infty$.~\cite{takahashi99} 
There is also a corresponding finite-size shift in the excitations close to $k=\pi$, which is characteristic of the Heisenberg chain but not present in the model
in higher dimensions. The lowest triplet gap in the neighborhood of $k=\pi$, where we will use only $k=\pi+2\pi/L$, behaves for large $L$ as \cite{caux}
\begin{equation}
\Delta(\pi+2\pi/L) = \Delta(\pi) + c(L)\frac{2\pi}{L},
\label{umklapp}
\end{equation}
where $\Delta(\pi) \sim 1/L$ but with a multiplicative logarithmic correction.
In Fig~\ref{velexact1dpi} we show the behavior of the corresponding velocity estimate, 
\begin{equation}
c(L)=\frac{L}{2\pi}[\Delta(\pi+2\pi/L) - \Delta(\pi)],
\label{cldef}
\end{equation}
as a function of the inverse chain length. A smooth monotonic (asymptotically linear) approach to the known velocity, $c=\pi/2$ can be observed as $L \rightarrow \infty$.
There are no signs of any remaining log corrections as a function of $1/L$ in this estimate. Our results are consistent with a remaining linear finite-size correction.

\begin{figure}
\center{\includegraphics[width=7cm, clip]{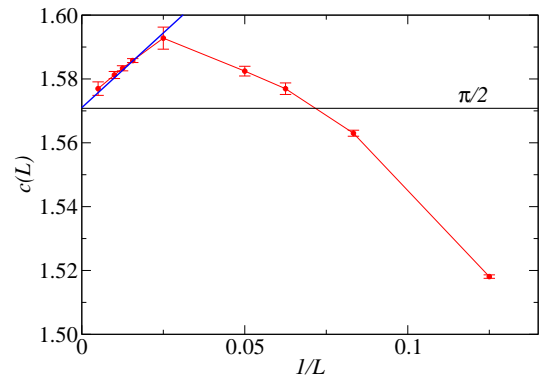}}
\vskip-2mm
\caption{(Color online) Velocity estimate for the Heisenberg chain from the triplet gap close to $k=0$; $c(L) = \Delta(k)/k$, where $k=2\pi/L$, shown as a function 
of the inverse system size. The known velocity $c=\pi/2$ in the thermodynamic limit is indicated by the horizontal line. The line through the larger system sizes is not a fit but is
drawn to match the known velocity and the data for those system sizes.}
\label{velexact1d}
\vskip-4mm
\end{figure}

The velocity  of the spinons can also be extracted from the lowest triplet gap at $k=2\pi/L$ by using the simple
estimator $L{\Delta(2\pi/L)}/{(2\pi)} \to c$ as $L \rightarrow \infty$.  Results  are shown in Fig~\ref{velexact1d}.
Also in this case we observe the estimate approaching the correct value of $c$ as $L \rightarrow \infty$, but with a non-monotonic 
behavior with a maximum for $L \approx 40$ before an apparently linear asymptotic approach to the correct value. It should be noted that the spectral
weight (before normalizing the correlation function) vanishes as $k \to 0$, which implies that the $k=2\pi/L$ correlation function 
computed in the QMC simulations becomes very noisy for large system sizes. It is worth noting here that, because of the non-monotonic behavior
seen in Fig~\ref{velexact1d}, a calculation using exact (numerical) diagonalization of the Hamiltonian would appear inconsistent with the known velocity,
because large enough system sizes required to go beyond the maximum at $L \approx 40$ cannot be reached.

\subsection{Exponential-form projection}
\label{sec:ct}

In this section we will discuss a different way of extracting the gap, using a systematic way to analyze moments of the spectrum based 
on information contained in the imaginary-time correlations. Such a scheme was recently introduced within $T >0$ QMC calculations,\cite{SuwaT2014}
and we here present a generalization to projector QMC. In addition, we discuss improvements in the extrapolations required to obtain
unbiased results.

To make the connection with the previous version of the method more transparent we will here use the exponential-form projection with 
continuous imaginary time. The ground state is again projected out of a trial state in the valence-bond basis, but now using $e^{-\beta H}$ instead
of $(-H)^m$. Taylor expanding the exponential (as was also done previously in projector QMC, e.g., in Ref.~\onlinecite{farhi12}), the 
scheme then closely resembles $T>0$ SSE QMC algorithms.\cite{sandvik99} The normalization, which replaces the partition function in $T>0$ methods, 
is expressed as 
\begin{eqnarray}
Z' & = & \langle \psi_t | e^{-\beta H} | \psi_t \rangle \nonumber \\
   & =  &\langle \psi_t | T_{\tau} {\rm exp}\left( - \int_0^{\beta} d\tau H(\tau) \right) | \psi_t \rangle \label{eqn:ct} \\
   & = & \langle \psi_t | 1+ \sum_{m=1}^{\infty} (-1)^m \int_0^{\beta} d\tau_1 \int_{\tau_1}^{\beta} d\tau_2 \cdots \nonumber \\ 
    &&~~~\cdots \int_{\tau_{m-1}}^{\beta} d\tau_m \prod_{i=1}^m H(\tau_i) | \psi_t \rangle \nonumber
\end{eqnarray}
where $T_{\tau}$ indicates time ordering and $H(\tau)$ is a Hamiltonian acting at imaginary time $\tau$. If the Hamiltonian
$H(\tau)$ here is actually time dependent, we obtain the non-equilibrium QMC scheme developed in Ref.~\onlinecite{degrandi11}.
Here $H$ is time-independent, and
we formally use the time-dependent formalism only as a convenient way of accessing imaginary time in the configuration space directly. In
principle, we can also use the time dependence corresponding to the interaction representation,\cite{beard96,prokofev98,sandvik97b} where
only the $x$- and $y-$ parts of the interaction appears in the operator product, but here we stay close to the formulation also used
in the SSE method and in the $H$-power method discussed above, and expand with the full Hamiltonian. Note that the $H$-power method
is exactly recovered if the time integrals are completed, and one can in practice also equivalently use the power method and just generate 
the ordered time sequences at random,\cite{sandvik97b} or generate them in the process of the updates as in the world-line continuous time
algorithm.\cite{beard96}

\subsubsection{Generalized Moment Method}
\label{sec:moments}

Our approach introduced here to extract the gap is a generalization to projection QMC of the moment method proposed recently.\cite{SuwaT2014} The 
Fourier-transformed dynamical correlation is exploited to derive a sequence of gap estimators. We will first explain how the moment of the 
dynamical correlation function is related to the gap, and then derive increasingly precise gap estimators. Finally, we will 
show how these estimates are extrapolated to the limit in which the exact gap is recovered.

We will use the following dynamical correlation function computed with the exponential projector:
\begin{eqnarray}
C(\tau, \tau_0) &=& \frac{1}{Z'} \langle \psi_t | e^{ - ( \beta - \tau - \tau_0 ) H} A^{\dagger} e^{-\tau H} A e^{- \tau_0 H} | \psi_t \rangle \nonumber \\
&=& \frac{1}{Z'} \sum_{\ell, p, q} b_{\ell, p, q} \, e^{-\beta E_p} e^{- \tau ( E_{\ell} - E_p )} e^{ - \tau_0( E_q - E_p )} \nonumber \\
&\rightarrow& \sum_{ \ell \geq 1 } b_{\ell} e^{-\tau \Delta_{\ell} } \quad ( \beta, \tau_0 \rightarrow \infty ) \label{eqn:c-tau},
\end{eqnarray}
where we use the definitions
\begin{eqnarray}
b_{\ell,p,q} & = & \bar{c}_p c_q \langle p | A^{\dagger} | \ell \rangle \langle \ell | A | q \rangle, \\
b_{\ell} & \equiv & b_{\ell,0,0} = | c_0 \langle \ell | A | 0 \rangle |^2, \\
\Delta_{\ell} & = &  E_{\ell} - E_0.
\end{eqnarray}
We assume that $A | 0 \rangle \neq 0 $, $\langle 0 | A | 0 \rangle = 0$, and $\Delta_1 > 0$, all of which apply here.

Let us first consider the moment of the asymptotic correlation function in the limit $\beta, \tau_0 \rightarrow \infty$:
\begin{eqnarray}
I_n^{\infty} &=& \int_0^{\infty} d \tau \tau^nC(\tau, \tau_0) \nonumber \\
&=&  \int_0^{\infty} d \tau \, \tau^n \sum_{ \ell \geq 1 } b_{\ell} \, e^{-\tau \Delta_{\ell} } \nonumber \\
&=& \sum_{\ell \geq 1 } \frac{ b_{\ell} }{ \Delta_{\ell}^{n+1} } n! \nonumber \\
&\sim& \quad \frac{b_1}{\Delta_1^{n+1}} n! \quad ( n \gg 1 ). \label{eqn:I_n_inf}
\end{eqnarray}
Then the lowest gap can, in principle, be obtained using an appropriate ratio, e.g.,
\begin{align}
\frac{(n+1) I_n^{\infty}}{ I_{n+1}^{\infty} } \rightarrow \Delta_1 \quad ( n \rightarrow \infty ) .
\label{eqn:moment2gap}
\end{align}
In practice, however, the projection time $\beta$ cannot be infinite in simulations, and we have to consider carefully the effects of
finite $\beta$. In addition, the range of integration, $\tau \in (0,\beta_{\rm int})$, is different from (less than) $\beta$ because the
imaginary-time correlations are measured from the reference point $\tau_0=\beta/2$ at the center of the projected trial state [noting that
Eq.~(\ref{eqn:ct}) corresponds to projection of the bra and ket state with ${\rm e}^{-\beta/2}$], or, alternatively, between two points located symmetrically
within the projection range $\tau \in (0,\beta_{\rm int})$, and one has to stay well away from the boundaries (trial states) for unbiased measurements.
Moreover, in many cases the statistical errors grow too large at large times, as mentioned previously. Thus, in practice $\beta_{\rm int} < \beta/2$.

The moments for finite $\beta$ and $\beta_{\rm int}\leq \beta/2$ take the form
\begin{align}
I_n &= \int_0^{\beta_{\rm int}} d \tau \, \tau^n C(\tau, \tau_0) \nonumber \\
&= \frac{1}{Z'} \int_0^{\beta_{\rm int}} d \tau \, \tau^n \sum_{\ell, p} d_{\ell, p} e^{- \tau \Delta_{\ell, p}} \nonumber \\
&= \frac{1}{Z'} \sum_{\ell, p} \frac{d_{\ell, p} \, n!}{ \Delta_{\ell, p}^{n+1} } \left( 1 - \sum_{m=0}^n P(m) \right),
\end{align}
where 
\begin{eqnarray}
d_{\ell, p} & = & e^{- \beta_{\rm int} E_p} \sum_q b_{\ell, p, q} e^{- \tau_0 ( E_q - E_p ) },\\
\Delta_{\ell, p} & = & E_{\ell} - E_p, 
\end{eqnarray}
and
\begin{equation}
P(m)= \frac{ (\beta_{\rm int} \Delta_{\ell, p} )^m e^{ - \beta_{\rm int} \Delta_{\ell, p} } }{ m! }
\end{equation}
is a properly normalized Poisson distribution;
\begin{equation}
 \sum_{m=0}^{\infty} P(m) = 1.  
\end{equation}
Owing to the finite integration range, the Poisson term does not vanish. As a consequence, the ratio of the moments for finite $\beta$ 
does not contain information of the gap in the limit $n \rightarrow \infty$. Instead, we have a completely different limiting behavior,
\begin{equation}
\frac{ I_n }{ I_{n+1} } \sim \frac{ n + 2 }{ \beta_{\rm int} ( n + 1 ) } \rightarrow \frac{1}{\beta_{\rm int} }  \qquad ( n \rightarrow \infty) ,
\end{equation}
independent of the gap.
We can overcome this difficulty and devise a proper gap estimator by using the Fourier transformation. Here we express the
moments in a different way using the following expansion:
\begin{align}
 \frac{ I_{2n}^{\infty} }{ ( 2n )! } &= \lim_{\beta_{\rm int}, \tau_0 \rightarrow \infty} \frac{ ( -1 )^n }{ \omega_1^{ 2n } } \sum_{k=0}^n x_{n,k,0} R( \omega_k ) \label{eqn:even-moments} \\
 \frac{ I_{2n-1}^{\infty} }{ ( 2n - 1 )! } & = \lim_{\beta_{\rm int}, \tau_0 \rightarrow \infty} \frac{ ( -1 )^{n-1} }{ \omega_1^{ 2(n-1) } } \sum_{k=1}^n x_{n,k,1} \frac{ J( \omega_k ) }{ \omega_k } , \label{eqn:odd-moments}
\end{align}
where $R(\omega_k)$ and $J(\omega_k)$ are the real and imaginary parts, respectively, of the Fourier-transformed correlation function, i.e.,
\begin{align}
\int_0^{\beta_{\rm int} } d\tau \, e^{i \tau \omega_k} C(\tau, \tau_0) = R(\omega_k) + i J(\omega_k) \label{eqn:Fourier-corr},
\end{align}
where $\omega_k = 2 \pi k / \beta_{\rm int} $ $(k \in \mathbf{Z})$, and the key coefficients are $x_{1,1,1}=1$ and 
\begin{align}
x_{n,k,m}= \frac{ 1 }{ \displaystyle \prod_{m \leq j \leq n, j \neq k} ( k + j )( k - j ) }.
\end{align}
When deriving these equations we have considered the expansion in $\omega_k$ on the right-hand side of Eq.~(\ref{eqn:even-moments}) and Eq.~(\ref{eqn:odd-moments}), where the
lowest orders then cancel. The coefficients $x_{n,k,m}$ can be solved for by using the inverse of the Vandermonde matrix.\cite{Neagoe1996} 

Combining the results above, we obtain the following improved gap estimator:
\begin{align}
\hat{\Delta}_{( n,\beta_{\rm int} )} = - \omega_1^2 \, \frac{ \displaystyle \sum_{k=1}^n x_{n,k,1} \frac{ \displaystyle J(\omega_k) }{ \omega_k } }{ \displaystyle \sum_{k=0}^n x_{n,k,0} R(\omega_k) } .
\label{eqn:gap_Fourier}
\end{align}
Remarkably, this estimator is asymptotically unbiased and the limits are interchangeable (see Appendix for the analytical derivation):
\begin{align}
 \Delta_1 & = \lim_{n \rightarrow \infty} \lim_{\beta_{\rm int}, \tau_0 \rightarrow \infty} \hat{\Delta}_{ (n,\beta_{\rm int}) } \nonumber \\ 
& = \lim_{\beta_{\rm int}, \tau_0 \rightarrow \infty} \lim_{n \rightarrow \infty} \hat{\Delta}_{ (n,\beta_{\rm int}) }. \label{eqn:gap_Fourier_limits}
\end{align}
Due to the commuting limits, observing convergence of the estimator~(\ref{eqn:gap_Fourier}) in large $n$ and $\beta_{\rm int}$ taken in any
convenient fashion will deliver an unbiased result for the gap.

In principle there is also a dependence on the  $\beta$ value used in the exponential projection, in addition to the dependence on $n$
and $\beta_{\rm int}$. Above we have assumed that $\beta$ is sufficiently large for quantities computed by integration up to $\beta_{\rm int}$
have converged to the $\beta \to \infty$ limit, and in the QMC simulations we also monitor this convergence. Naturally, the value of $\beta$ 
required grows with $\beta_{\rm int}$.

\begin{figure}
\center{\includegraphics[width=8cm, clip]{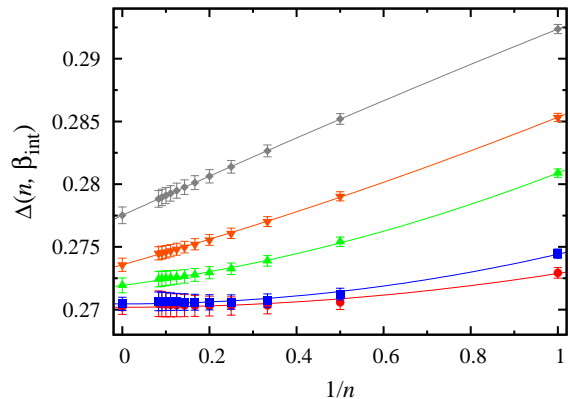}}
\vskip-2mm
\caption{(Color online) Extrapolation of $\Delta(n \rightarrow \infty, \beta_{\rm int})$ for $\beta_{\rm int}=6$ (diamonds), $7.68$ (lower triangles), 
$9.6$ (upper triangles), $19.2$ (squares), and $32$ (circles). A quadratic polynomial in $1/n$ was fitted to the data points for each $\beta_{\rm int}$ 
and the extrapolated $n \to \infty$ value with error bar is shown in each case.}
\label{fig:gap-n}
\vskip-4mm
\end{figure}

\subsubsection{Performance Check for the Heisenberg Chain}

We have checked the validity of the estimator~(\ref{eqn:gap_Fourier}) for the Heisenberg chain of 16 spins with periodic boundary conditions. 
The projection length $\beta$ is set large enough to see the asymptotic behavior of the dynamic correlations on the imaginary-time axis. The correlations were 
measured from the center of the QMC configuration; that is, from $\tau_0 = \beta / 2$. As in the previous section, the loop algorithm and improved 
correlation-function estimator \cite{sandvik10a} are used. We consider the triplet gap at $k=\pi$ and the operator $A$ in Eq.~(\ref{eqn:c-tau}) is then explicitly given by
$A=\sum_r \mathbf{S}_r (-1)^r$ and 
\begin{equation}
A^{\dagger}e^{\tau H} A = \sum_{\alpha=x,y,z} \sum_{r, r'} S_r^{\alpha} e^{\tau H} S_{r'}^{\alpha} (-1)^{r-r'}.
\end{equation}
The Fourier-transformed correlation functions~(\ref{eqn:Fourier-corr}) were directly measured in the simulation, and no discretization error is
introduced. The gap estimators~(\ref{eqn:gap_Fourier}) for several $n$ and $\beta_{\rm int}$ were calculated and the errors estimated by the jackknife 
analysis~\cite{Berg2004} (bootstrapping would produce statistically equivalent result).

As shown above, our estimator (\ref{eqn:gap_Fourier}) converges to the exact gap in the limit of infinite $n$ and $\beta_{\rm int}$. In practice, if good convergence 
is observed within statistical errors, a gap value obtained from a sufficiently large $n$ and $\beta_{\rm int}$ within the converged range can be used as the final 
estimation, or some appropriate functional form can be used for extrapolation. However, an important issue here is that the statistical errors grow with the two parameters. 
We therefore inevitably encounter a trade-off problem between systematical and statistical extrapolation errors. Here we will demonstrate that the statistical error 
can be optimized by proper extrapolations while keeping the estimation unbiased. In the procedure used below, the extrapolation is taken for $n \to \infty$ first and then
for $\beta_{\rm int} \to \infty$, though, as discussed above, other ways to accomplish the limit are also possible. 

\begin{figure}
\center{\includegraphics[width=8cm, clip]{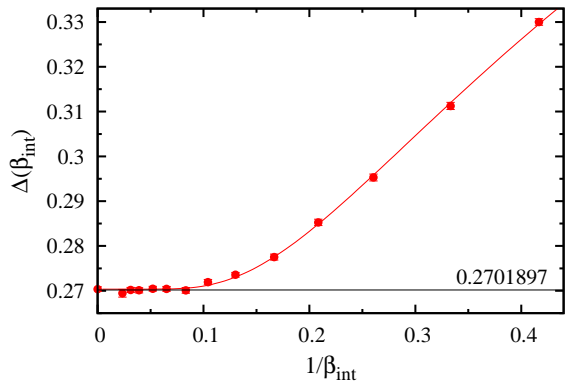}}
\vskip-4mm
\caption{(Color online) Extrapolation of the gap for $\beta_{\rm int} \rightarrow \infty$. An exponential function 
$f(\beta_{\rm int}) = \Delta + a \times \exp( - b \beta_{\rm int})$ (curve) has been fitted to the data points, where 
$a$ and $b$ are positive real numbers. The exact gap value calculated from the full diagonalization is shown by 
the horizontal line.}
\label{fig:gap-beta}
\vskip-2mm
\end{figure}

Examples, based on $2^{20}$ ($\sim 10^6$) Monte Carlo samples, of $n \rightarrow \infty$ extrapolations are shown in Fig.~\ref{fig:gap-n}. The leading 
finite-$n$ correction is linear in $1/n$ [in accord with Eq.~(\ref{eqn:n-inf}) and~(\ref{eqn:gap-fit}) in the Appendix], and we include also a quadratic 
term to obtain good fits to the data for the full $n$-range. The limit $\beta_{\rm int} \rightarrow \infty$ is taken next using the resulting values of 
the $n \rightarrow \infty$ extrapolations, as shown in Fig~\ref{fig:gap-beta}. Here an exponential function is used for the data fit (again, according to results 
derived in the Appendix) to extrapolate the final result for the gap. Though other ways of extrapolating to $n,\beta_{\rm int} \rightarrow \infty$ are possible, 
the above protocol is convenient because the extrapolation for $n$, which is taken first, is relatively easier than that for $\beta_{\rm int}$.

\begin{figure}
\center{\includegraphics[width=7cm,clip]{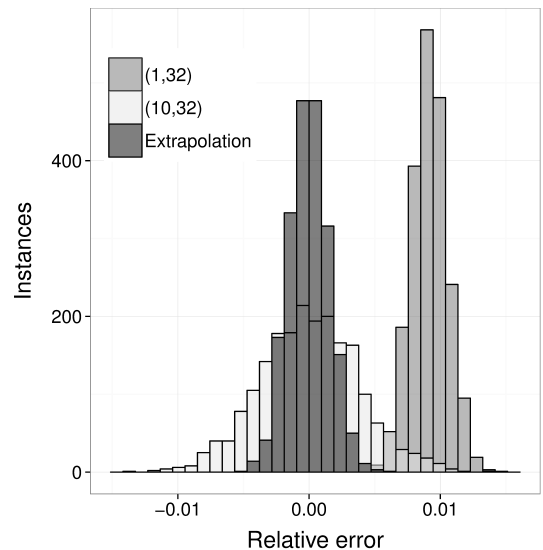}}
\vskip-2mm
\caption{Histograms of the density of the relative error of the $(n\to \infty,\beta_{\rm int}\to \infty)$ extrapolated gap, along with those based on 
the estimator~(\ref{eqn:gap_Fourier}) for fixed $(n,\beta_{\rm int})=(1, 32)$ and $(10, 32)$. The error was calculated as $( \hat{\Delta} - \Delta ) / \Delta$, 
where $\hat{\Delta}$ is the generalized-moment estimated value from $2^{20}$ ($\sim 10^6$) Monte Carlo samples and $\Delta$ is the exact value for the 
$L=16$ Heisenberg chain. The histograms include results of 2048 independent simulations.}
\label{fig:gap-hist}
\vskip-2mm
\end{figure}

As a test of the unbiased nature of the extrapolation scheme, distributions of the relative gap error are shown in Fig.~\ref{fig:gap-hist}. Histograms
were collected based on 2048 independent simulations of the $L=16$ Heisenberg chain. Results based on the extrapolation procedure discussed above are
compared with those of individual gap estimators for $(n,\beta_{\rm int}) = (1, 32)$ and $(10, 32)$. The $(1,32)$ estimator clearly has a non-zero systematic 
error remaining, which is similar to (but smaller than) the conventional second moment estimator.\cite{SuwaT2014} The $(10,32)$ estimator has a small enough 
systematical error (the histogram being centered very close to zero) but has a large statistical error (wide distribution). The extrapolated estimation is unbiased 
and has a small statistical error.

\section{The velocity from winding number fluctuations}
\label{sec:winding}

We here discuss how to use winding numbers to compute the velocity. This method has been known for some time,\cite{kaul08} and was recently 
applied to the 2D XY model \cite{jiang11b} and the 2D Heisenberg model.\cite{jiang11a} We begin by briefly recollecting some key aspects about winding numbers 
in finite-temperature QMC simulations, in particular the SSE method we use for these calculations. We then present results for the Heisenberg chain.

\subsection{Winding numbers in QMC simulations} 

QMC simulations at finite temperature are based on some mapping of the partition function of a quantum system in $D$ dimensions to an effectively 
equivalent $(D+1)$-dimensional classical system, where the additional dimension of the configurations corresponds to imaginary (Euclidean) time.
The effective length of the system in the time dimension is $c\beta$, where $\beta=1/T$ (setting $k_B=1$) and the configurations are time-periodic. For 
a system with conserved particle number, which in the case of spin model corresponds to conserved magnetization along the quantization ($z$) axis, 
imposing periodic boundary in the spatial dimensions leads to another, topological number associated with the configurations---the winding number
representing permutations when particles are propagated once or multiple times around the periodic system. The winding numbers were first used
in QMC calculations of the superfluid stiffness of interacting bosons.\cite{pollock87}

The SSE QMC algorithm,\cite{sandvik91,sandvik92,sandvik97} is based on a Taylor expansion of the imaginary-time evolution operator (the Boltzmann operator) 
${\rm e}^{-\beta H}$, and similarities with the projector approach discussed in the previous section were already pointed out.
Each SSE configuration is associated with some power $H^n$ of the Hamiltonian propagating a basis state (here in the standard computational basis of $z$ 
spin components), and these powers are sampled stochastically to all contributing orders. The trace over all basis state is also sampled. The average 
expansion power $\langle n\rangle$ in this procedure is proportional to $\beta$; $\langle n\rangle = \beta \langle H\rangle \propto \beta N$. In simulations, 
the state propagation is broken up into individual paths corresponding to strings of $n$ of the individual local terms of the Hamiltonian, forming successions 
of $n$ evolving basis states, similar to those in path integrals. For a Heisenberg model the Hamiltonian terms are the diagonal operators $S^z_iS^z_j$ (in 
practice with a constant added) and off-diagonal $S^+_iS^-_j+S^-_iS^+_j$ operators, the latter of which transport spin and are associated with currents 
$J_a = \pm 1$ in the lattice direction $a$ corresponding to the site-pair $i,j$. The winding number in the $a$ lattice direction is defined in terms of 
the currents as
\begin{equation}
W_{a} = \frac{1}{L_a}\sum_{p=1}^n J_{a}(p),
\end{equation}
where the index $p$ corresponds to the location of the transport ``event'' in the string of $n$ operators and $L_a$ is the length of the lattice in 
the $a$ direction. Defined in this way, the winding numbers are integers. It should be noted that the definition of the winding number is exactly 
the same in SSE and world-line methods \cite{suzuki77,hirsch82} and one can also think of the SSE configurations as consisting of world lines (for up 
and down spins in the case of $S=1/2$ quantum spin systems).

The spatial winding number $W_r$ measures the net spin transported around the periodic lattice in the $r$ direction in the course of the periodic propagation 
in imaginary time. Equivalently, this is the number of world lines (up ones minus down ones divided by two) crossing through a plane drawn along the time axis 
perpendicular to the $r$ axis. Since the total $z$ magnetization $M_z$ is conserved, one can also think of $M_z$ as a winding number; the net number of world 
lines crossing a plane drawn at an arbitrary time point perpendicular to the time axis, which is just the magnetization computed in the stored basis state;
\begin{equation}
W_\tau = M_z =\sum_{i=1}^N S^z_i.
\end{equation}
The expectation values of the squared winding numbers (i.e., the winding number fluctuations) are related to two important thermodynamic quantities; 
the spin stiffness,
\begin{equation}
\rho_s = \frac{1}{2\beta}\left ( \left \langle W_x^2 \right \rangle  + \left \langle W_y^2 \right \rangle  \right ),
\label{rhos}
\end{equation}
and the uniform magnetic susceptibility,
\begin{equation}
\chi= \frac{\beta}{N} \left \langle M_z^2 \right \rangle = \frac{\beta}{N}\left \langle W_\tau^2 \right \rangle.
\label{ususc}
\end{equation}
The technicalities of implementing these observables in SSE calculations have been discussed extensively in the literature; see Ref.~\onlinecite{sandvik10b}
for a recent review.

\subsection{The cubic criterion and the velocity}

In the high-temperature limit $T\to \infty$, the magnetization fluctuations of any system are maximized and therefore $\langle W_\tau^2 \rangle > 0$  according to 
Eq.~(\ref{ususc}). For an unfrustrated antiferromagnet the ground state is a singlet, and, on account of the presence of a singlet-triplet finite-size gap, 
$\langle W_\tau^2 \rangle \to 0$ when $T\to 0$ for any finite system. In contrast to these limits of $\langle W_\tau^2 \rangle$, for the spatial winding number in 
direction $r$ ($r=x,y,\ldots$) we have $\langle W_r^2 \rangle \to 0$ when $T\to \infty$ on account of there being no quantum fluctuations when the imaginary-time length $\beta \to 0$ and there are no 
contributions from expansion powers $n>0$ (in the case of SSE---in world-line methods there will similarly be no transport events causing shifts of the world lines).
In the limit $T\to 0$, for a system with long-range order (or a "quasi-ordered" 1D system with power-law decaying correlations), the stiffness constant converges to a 
non-zero value for any $L$, and according to Eq.~(\ref{rhos}) we must then have a divergence $\langle W_r^2\rangle \sim 1/T$. These different behaviors of the spatial 
and temporal winding numbers versus temperature guarantees that there is a crossing point,
\begin{equation}
\left\langle W_r^2(\beta^*) \right\rangle = \left \langle W_\tau^2 (\beta^*) \right\rangle,
\label{cubic}
\end{equation}
at some unique value of $\beta=\beta^*(L)$ for given size $L$.

The winding numbers characterize global fluctuations of the system in the different spatial and temporal directions. It is then natural to define a system as having 
cubic space-time geometry when Eq.~(\ref{cubic}) holds (with the lattice length $L$ the same in all spatial directions). The aspect ratio $L/\beta^*$ should then
directly correspond to the velocity of the long-wave-length excitations. In some cases this can be shown directly based on low-energy field theory 
\cite{kaul08,jiang11a,jiang11b} but even in the absence of such descriptions the arguments are very general and one can expect the conclusion $c=L/\beta^*$ 
to always hold for a system with linear dispersion, though we are not aware of any formal proofs in the general case.

\subsection{Test on the Heisenberg chain}

\begin{figure}
\center{\includegraphics[width=8cm, clip]{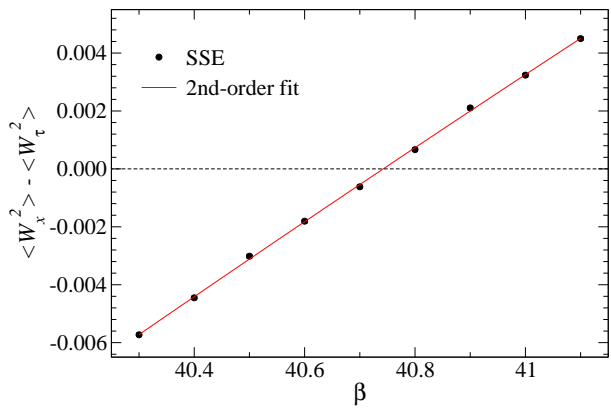}}
\vskip-2mm
\caption{(Color online) Difference between the spatial and temporal winding numbers versus the inverse temperature in
simulations of a $64$-site Heisenberg chain. A second-order polynomial has been fitted to the data points, and this curve
is used to determine the value $\beta=\beta^*$ at which cubic condition $\langle W_x^2\rangle=\langle W_\tau^2\rangle$ holds
(i.e., intersection with the horizontal dashed line).}
\label{wfit_1d}
\vskip-2mm
\end{figure}

To find the point where the cubic criterion $\langle W_x^2\rangle=\langle W_\tau^2\rangle$ is satisfied, we simulate a system at
several values of $\beta$ in the region where $\langle W_x^2\rangle \approx \langle W_\tau^2\rangle$ based on initial explorations
and knowledge of the approximate value of the velocity. We fit a low-order polynomial (typically second- or cubic-order) to the difference 
$\langle W_x^2\rangle-\langle W_\tau^2\rangle$ and solve the resulting equation for the $\beta$-value for which the cubic criterion is satisfied. 
This procedure is illustrated in Fig.~\ref{wfit_1d} for a Heisenberg chain with $L=64$ spins. From this procedure we obtain $c(L) = L/\beta^*(L)$, 
which can be extrapolated to $L \rightarrow \infty$.  With independent data points, the statistical error of $c(L)$ at fixed $L$ can be determined by repeating 
the procedure multiple times with added Gaussian noise of standard deviation equal to the error bars of the data points, whence the
standard deviation of the extracted crossing point is the error bar.

\begin{figure}
\center{\includegraphics[width=8cm, clip]{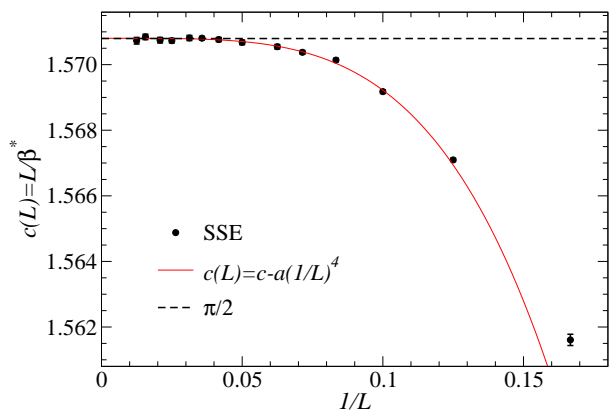}}
\vskip-2mm
\caption{(Color online) Size dependence of the velocity obtained with the cubic-criterion for the Heisenberg chain,
along with a fit corresponding to a leading $\sim 1/L^4$ size correction. The dashed line is at the rigorously known velocity
$c=\pi/2$.}
\label{cwind_1d}
\vskip-2mm
\end{figure}

Fig.~\ref{cwind_1d} shows results of such a procedure for several chain lengths $L$, graphed versus $1/L$. We are not aware of any theoretical predictions
for the size dependence of this definition of $c(L)$,  but the data for the larger systems are well described by a constant (the final infinite-size value of $c$) 
plus a term proportional to $1/L^4$. Using this fitting form leads to a value of $c$ completely consistent with the known value $c=\pi/2$, as is clear from
Fig.~\ref{cwind_1d}.

\section{Hydrodynamic Relationship}
\label{sec:hydromethod}

A well known way to extract the velocity in an antiferromagnet is to use the analogue of a hydrodynamic relationship between the
velocity, the spin stiffness (helicity modulus), and the transversal susceptibility (which is the analogue of a mass density):\cite{halperin69}
\begin{equation}
c= \sqrt{\frac{\rho_s}{\chi_\perp}}.
\label{chydro1}
\end{equation}
In a QMC calculation in which the spin-rotation symmetry is not explicitly broken, one cannot compute ${\chi_\perp}$ directly, but one can
use the fact that the rotationally-averaged uniform susceptibility $\chi$ computed in the $z$ basis, as in Eq.~(\ref{ususc}), becomes $2/3$ 
of a transversal (e.g., $x$) component when $T \to 0$ in the thermodynamic limit (since the longitudinal component vanishes at $T=0$). Thus, 
one can obtain ${\chi_\perp}$ as $(3/2)\chi$ by taking the limit $L \to \infty$ before the $T\to 0$ limit, while taking the limits in the opposite 
order does not work because then $\chi \to 0$ due to the finite-size gap between the $M_z=0$ and $M_z>0$ magnetization sectors. In contrast, 
for $\rho_s$ the limit $T\to 0$ has to be taken before $L \to \infty$, because the system in the thermodynamic limit only has stiffness (N\'eel order) 
exactly at $T=0$. In this case, too, a factor $3/2$ has to be included in the finite-size estimate to account for rotational averaging of 
the relevant transversal components.

The procedure to obtain $\rho_s$ in the thermodynamic limit is relatively straight-forward with an extrapolation of $\rho_s(L,T\to 0)$ using
a polynomial in $1/L$, while the extrapolations requiring $L \to \infty$ first in the $\chi$ calculation is more cumbersome. Results obtained 
for $c$ in this way \cite{sandvik10b} have large error bars compared to the result of the winding number method presented above in Sec.~\ref{sec:winding2d}.

In order to define a finite-size velocity estimate $c(L)$ based on Eq.~(\ref{chydro1}) directly in the $T=0$ limit we here use a modification of
the relationship. We use the susceptibility at finite momentum ${\bf q}$,
\begin{equation}
\chi({\bf q}) = \frac{1}{N} \int_0^\beta \langle M_z(-{\bf q},\tau)M_z({\bf q},0)\rangle,
\label{xqdef}
\end{equation}
where the magnetization at non-zero momentum is the same as $A$ in Eq.~(\ref{akszdef}) but without the normalization by $N^{-1/2}$.
We can use the smallest momentum $q=2\pi/L$ for a given system size and approach the $q=0$ limit as $L \to \infty$. If we here define
$\rho_s(L)$ without the rotational factor $3/2$ discussed above, then the same rotational factor should also not be used in the susceptibility,
and we define the velocity for finite size as
\begin{equation}
c(L)= \sqrt{\frac{\rho_s(L)}{\chi(q=2\pi/L)}}.
\label{chydro2}
\end{equation}
We will compute this quantity using SSE simulations at sufficiently large $\beta$ to achieve ground state convergence for
each $L$ studied. 

\subsection{Test on the Heisenberg chain}

The hydrodynamic relationship (\ref{chydro1}) is normally applied in the magnetically ordered N\'eel state and one may then question its use in
a critical system. In the case of the Heisenberg chain we can rigorously see that it is a valid way to extract $c$. Since there is no long-range 
order, there is no distinction between longitudinal and transversal modes, but Eq.~(\ref{chydro1}) defined with rotationally averaged quantities 
should remain valid. According to the exact Bethe Ansatz solution, in the thermodynamic limit we have, in the units with $J=1$ and all
relevant constants set to $1$, $\rho_s=1/4$ (derived in Ref.~\onlinecite{hamer87}) and $\chi (q\to 0)=1/\pi^2$ (see., e.g., Ref.~\onlinecite{eggert94}) 
and, thus, according to Eq.~(\ref{chydro2}), we obtain the correct result $c=\pi/2$.

\begin{figure}
\center{\includegraphics[width=8cm, clip]{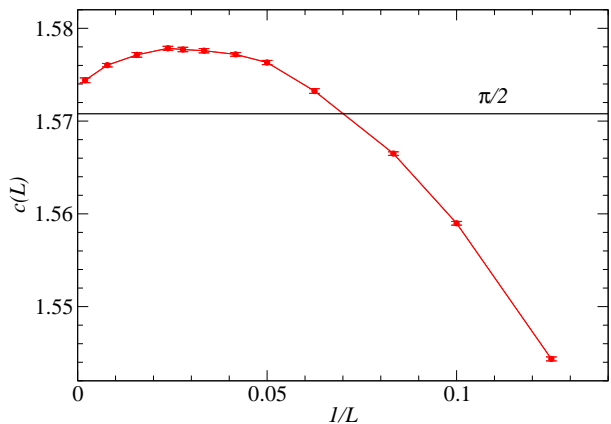}}
\vskip-2mm
\caption{(Color online) Size dependence of the velocity of excitations in the Heisenberg chain, defined according to the hydrodynamic
relationship in the form (\ref{chydro2}). The apparent poor convergence to the exact result $c=\pi/2$ when $L \to \infty$ can be explained by
logarithmic corrections. The maximum can be related to the anomaly in the excitation spectrum at $k=2\pi/L$, which causes the maximum at the 
same $L$ in Fig.~\ref{velexact1d}.}
\label{hydro1D}
\vskip-2mm
\end{figure}

For finite system size, it is well known that both $\chi$ and $\rho_s$ exhibit strong logarithmic corrections to their infinite-size values which can be traced to the presence of marginal operators and it
is very difficult to extrapolate them; see, e.g., Ref.~\onlinecite{byrnes02} for $\rho_s$ and Ref.~\onlinecite{eggert94} for $\chi$ (where in the latter 
case the logarithmic correction in temperature is discussed but one should expect similar corrections for $q \to 0$ at $T=0$). The logarithmic corrections
do not cancel in the ratio of the two quantities in Eq.~(\ref{chydro2}), and as a consequence we also find that it is difficult to extrapolate $c(L)$ 
precisely to the thermodynamic limit. Results are shown in Fig.~\ref{hydro1D}. Although it is not possible to fully extrapolate to the exact results based 
on these data for system size up to $L=512$, the result for the largest size nevertheless deviates by only $0.2\%$ from the exact result. 

It is interesting to note that also this definition of $c(L)$ exhibits a non-monotonic behavior, with a maximum at $L \approx 40$, the same as in the
$c(L)$ value obtained previously from the gap at $k=2\pi/L$ in Fig.~\ref{velexact1d}. In the latter case, the maximum corresponds to an elevated excitation energy at
$k=2\pi/L$, which one should expect to lead to a reduction in $\chi(2\pi/L)$ because this quantity is given by a sum rule of $S(k,\omega)/\omega$.
By this sum-rule, in a single-mode approximation a larger $\omega_k$ would lead to a smaller $\chi(k)$, and even beyond the single-mode
approximation one should expect such an effect because the dominant spectral weight is at the gap.  If $\rho_s$ is not appreciably affected by this
finite-size anomaly, then indeed an elevated gap and reduced $\chi(2\pi/L)$ in Eq.~(\ref{chydro2}) can explain the maximum in Fig.~\ref{hydro1D}.

In light of the logarithmic corrections to $c(L)$ defined according to Eq.~(\ref{chydro1}), the apparent complete lack of any challenging corrections or non-monotonic 
behavior in the definition based on the cubic criterion for winding numbers (Fig.~\ref{cwind_1d}) is even more remarkable, since also this estimate involves quantities directly 
related to the susceptibility (temporal winding number) and spin stiffness (spatial winding number). There are also no apparent logarithmic corrections in the 
velocities defined based on gaps in Figs.~\ref{velexact1dpi} and \ref{velexact1d}. In Fig.~\ref{velexact1dpi}, which is based on the triplet gap close 
to $\pi$, the logarithmic corrections are avoided by subtracting the $k=\pi$ gap in Eq.~(\ref{umklapp}), while the gap close to $k=0$ used in 
Fig.~\ref{velexact1d} is not expected to be affected by logarithms. 
The latter gap at $k=2\pi/L$ scales as $ck + b/L^2$ for large $L$ according to our results in Fig.~\ref{velexact1d}, which can be explained from the known dispersion $\Delta_k = c\sin(k) = ck + dk^3 + \ldots$, if there
is a finite-size correction $\sim 1/L^2$ (due to irrelevant fields only).

\section{2D Heisenberg model}
\label{sec:2D}

The 2D spin-$1/2$ Heisenberg model on the square lattice spontaneously breaks spin rotation symmetry at $T=0$ in the thermodynamic limit. 
This leads to gapless linearly dispersing Goldstone modes in the vicinity of the momenta $(0,0)$ and $(\pi,\pi)$. The ground state in a finite periodic 
system is a total spin singlet. The long-range antiferromagnetic order in the thermodynamic limit is reflected in the energies of the $S>0$ 
quantum rotor states \cite{rotorrefs} which collapse onto the ground state as $\Delta_S \sim 1/L^2$ much faster than than the spin wave excitations, 
$\Delta \sim 1/L$. The rotor states, thus, become degenerate with the ground state as the system size increases, and have momenta $\mathbf{k}=(0,0)$ and $(\pi,\pi)$ 
for even and odd $S$ respectively. Combinations of rotor states with $S$ up to $\sim L$ can then be formed which are ground states with fixed direction of 
the N\'eel vector (the staggered magnetization), thus allowing for symmetry breaking in the thermodynamic limit. Here we provide an accurate determination
of the spin-wave velocity and also directly investigate the scaling of the rotor state.

\subsection{Velocity from winding numbers}
\label{sec:winding2d}

\begin{figure}
\center{\includegraphics[width=8cm, clip]{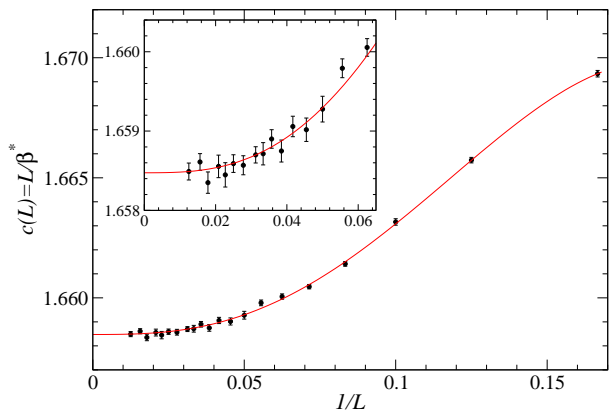}}
\vskip-2mm
\caption{(Color online) Size dependence of the velocity defined using the cubic-criterion for the 2D Heisenberg model, along with an 
fifth-order polynomial fit (including all system sizes $L \ge 6$) without linear and quadratic terms. The extrapolated velocity is $c=1.65847(4)$. 
The data for the largest system sizes are shown on a more detailed scale in the inset. The goodness of the fit is $\chi^2/{\rm dof} \approx 0.8$.}
\label{cwind_2d}
\vskip-2mm
\end{figure}

Since we expect the winding numbers with the cubic criterion to provide the best determination of $c$ we start with this approach.
Results based on the procedures discussed in the preceding section are shown in Fig.~\ref{cwind_2d} versus the inverse system length. 
Carrying out polynomial fits, we find that no linear and quadratic terms are required. Fourth- and fifth-order polynomials fits excluding
these terms and using all the $L \ge 6$ data give almost identical results for the $L \to \infty$ extrapolated velocity, with only the 
error bar somewhat larger for the fifth-order fit. The figure shows the fifth-order fit and the extrapolated velocity with it is $c=1.65847(4)$. 
While we do not know the physical reason for the leading cubic correction, we use it as the simplest empirical description of the data. 
Including also a quadratic term, it comes out equal to 0 within statistical errors and the extrapolated result does not change
appreciably. 

The above value of $c$ agrees within errors bars with the recent result using the same method by Jiang and Wiese,\cite{jiang11a} but our error 
bar is almost an order of magnitude smaller. We note that in Ref.~\onlinecite{jiang11a} no systematic extrapolation was carried out to the 
thermodynamic limit---instead an average was taken of results for system sizes in the range $L\in [24,64]$. Looking at the data in Fig.~\ref{cwind_2d} 
it is clear that, with the small error bars on the SSE data we have achieved here, an extrapolation is necessary to obtain a result with no remaining finite-size effects. 
To our knowledge, the above result is the most precise spin-wave velocity reported to date for the 2D Heisenberg model. Spin-wave theory with corrections up
to order $1/S^2$ gave $c=1.6638(3)$,\cite{canali92} where the uncertainty $3$ in the last digit reflects estimated numerical errors from evaluations of 
challenging integrals. Thus, to this order, the spin-wave result deviates by only $0.3\%$ from the correct value.

\subsection{Gap scaling of rotor states}

\begin{figure}
\center{\includegraphics[width=8cm, clip]{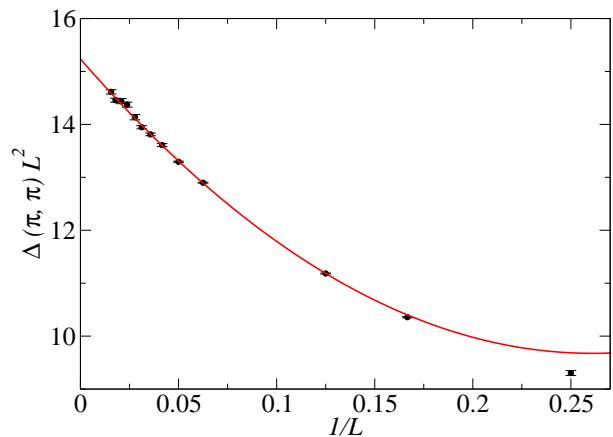}}
\vskip-2mm
\caption{(Color online) Size dependence of the lowest triplet gap $\Delta(\pi,\pi)$ multiplied by $L^2$ for the 2D Heisenberg model. This analysis
shows that the gap scales as $1/L^2$ for large $L$ and an extrapolation based on a polynomial in $1/L$ gives the estimate ${1}/{2\chi_{\perp}} = 7.62(2)$ 
for the uniform susceptibility.}
\label{rotor}
\vskip-2mm
\end{figure}

The quantum rotor excitation gap can be directly accessed by measuring the lowest triplet gap at $\mathbf{k}=(\pi,\pi)$. This energy scale is related to 
the uniform (transverse) magnetic susceptibility $\chi_{\perp}$ as:\cite{rotorrefs}
\be
E(S,L) = \frac{S(S+1)}{2L^2 \chi_{\perp}}. 
\ee 
From the behavior of the lowest triplet gap up to system sizes $L=64$, as shown in Fig.~\ref{rotor}, we indeed observe that $\Delta (\pi,\pi) \sim 1/L^2$ 
at large $L$, but there are also large corrections which we fit with additional higher-order powers of $1/L$. The extrapolation to infinite size
gives ${1}/{2 \chi_{\perp}} = 7.62(2)$. This is consistent with the value of susceptibility obtained using QMC calculations in small external 
magnetic fields to extract gaps between different spin sectors.\cite{Syljausen02}  

\subsection{Velocity from gaps}

\begin{figure}
\center{\includegraphics[width=7.5cm, clip]{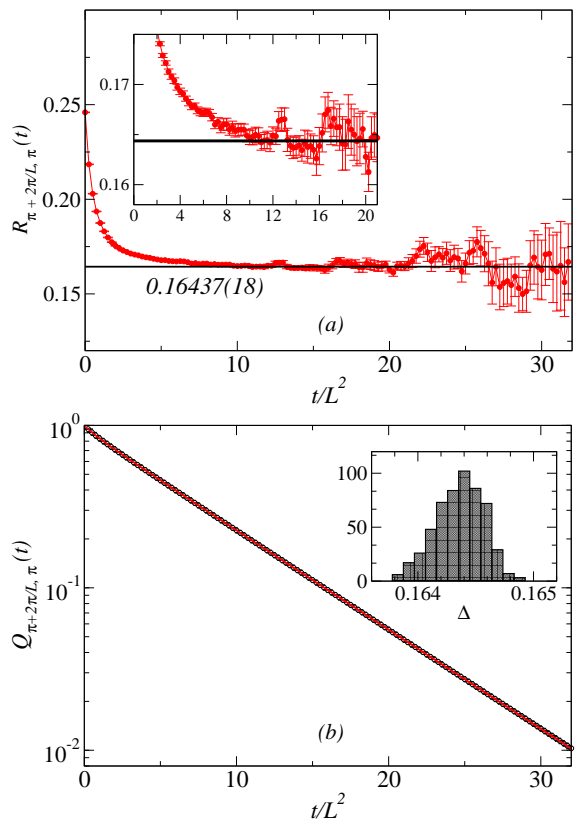}}
\vskip-2mm
\caption{(Color online) Extraction of the gap at $k=(\pi+2\pi/L,\pi)$ based on power-projected correlation functions, shown for the largest 
2D Heisenberg system studied; $L=64$. (a) The ratio $R_k(t)$, Eq.~(\ref{rdef}) converges to a finite gap value but the error bars are large for long imaginary 
times. Fitting $Q_k(t)$ to four $\delta$ functions according to Eq.~(\ref{gaps1}), as shown in (b), provides a more reliable gap estimation, as illustrated in 
the inset of (a). The thickness of the solid line here approximately represents the error bar. The inset of (b) shows the distribution of gap values obtained
from a large number of different bootstrap samples of the QMC data, from which the error bar of the gap estimate was computed.}
\label{neelL64}
\vskip-2mm
\end{figure}

The velocity of the spin waves can be estimated by measuring the triplet gap in the vicinity of $(\pi,\pi)$ and $(0,0)$. We here choose to measure the triplet gap at 
$\mathbf{k}_1=(\pi+2\pi/L,\pi)$ [or, equivalently, at $(\pi,\pi+2\pi/L)$, which we use for averaging] since the lowest triplet excitation energy is at $(\pi,\pi)$ and 
$\mathbf{k}_1$ is the closest allowed wave-vector to $(\pi,\pi)$ for a periodic system with linear size $L$. For this model we have carried out only $H$-power QMC
simulations. We illustrate the gap extraction with both the ratio method and the simplified spectral function with fitted $\delta$-functions in Fig.~\ref{neelL64}, using the
largest system sizes considered; $L=64$. Again, working with the spectral function produces much more stable results, though clearly the correlation ratio $R_k$ also 
converges to a constant consistent with the same gap. 

\begin{figure}
\center{\includegraphics[width=7.75cm, clip]{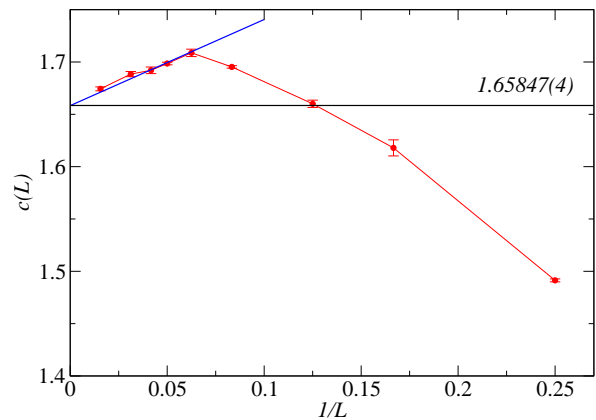}}
\vskip0mm
\caption{(Color online) The spin-wave velocity estimator $c(L)$ defined in Eq.~(\ref{cl2d})  shown for the 2D Heisenberg model as a function of $1/L$. 
The velocity for large $L$ is consistent with the value obtained from the winding numbers (Fig.~\ref{cwind_2d}), shown as a horizontal line. The 
line through the values for the larger system sizes is a guide to the eye.}
\label{neelgaps}
\vskip-2mm
\end{figure}

Since in the thermodynamic limit, the excitation energy of 
the spin waves equals $E (\mathbf{k}) = c |\mathbf{k}-(\pi,\pi)|$ in the vicinity of $(\pi,\pi)$, where $c$ is the spin-wave velocity, the estimator 
\be
c(L)=\frac{L}{2\pi}\Delta(\pi+ 2\pi/L,\pi)
\label{cl2d}
\ee
 should converge to $c$ as $L \rightarrow \infty$.  We graph this quantity in Fig.~\ref{neelgaps}, and it converges to the value of $c$ obtained above from the 
winding-number method when $L \rightarrow \infty$. Note that there is again a non-monotonicity as a function of $L$, similar to the case of the Heisenberg 
chain in Fig.~\ref{velexact1d}. In the latter case this behavior only was observed in the gap extracted close to $k=0$, however, while here we have used
the spectrum close to $(\pi,\pi)$.

\subsection{Hydrodynamic relationship}

\begin{figure}
\center{\includegraphics[width=7.75cm, clip]{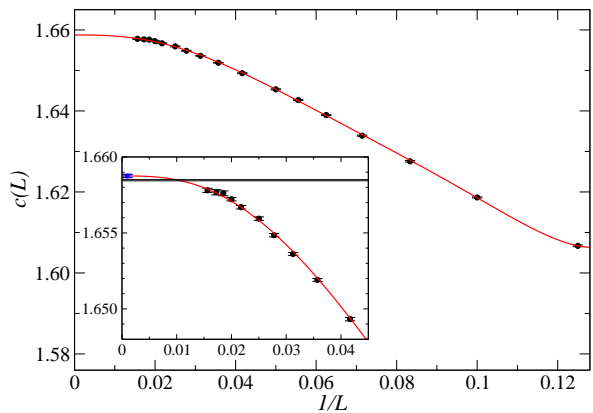}}
\vskip0mm
\caption{(Color online) Finite-size spin-wave velocity defined using the modified hydrodynamic relationship (\ref{chydro2}). The extrapolation
to infinite size is done using a sixth-order polynomial in $1/L$ without linear and quadratic terms (solid curve). The inset shows the data for the largest
systems on a more detailed scale. The point (blue) with error bar close to the $y$-axis indicates the statistical error of the $L \to \infty$ extrapolation. 
The barely separated horizontal lines shows the value of $c$ plus-minus one error bar obtained in Sec.~\ref{sec:winding2d} using the winding number method.}
\label{neelhydro}
\vskip-2mm
\end{figure}

We next consider the definition of $c(L)$ by the hydrodynamic relationship, Eq.~(\ref{chydro2}). We again go to sufficiently low temperature in SSE simulations
for any remaining finite-$T$ corrections to be negligible compared to the statistical errors, which typically meant $\beta=8L$. Results are graphed in 
Fig.~\ref{neelhydro}. Here again we expect the finite size corrections in the N\'eel state to be described asymptotically by a polynomial in $1/L$, 
but a rather high order of the polynomial is required to fit the data for the smaller system sizes. The behavior for the largest sizes again indicate (as 
in the case of the winding-number calculation discussed above) that the leading correction in $1/L$ is cubic. To get a statistically acceptable fit for all
$L \ge 8$ data we include terms up to order $6$, which gives $c=1.65875(10)$, which deviates by approximately $2.5$ error bars from the statistically much 
more precise value obtained in Sec.~\ref{sec:winding2d} based on the winding-number fluctuations. By comparing Figs.~\ref{cwind_2d} and \ref{neelhydro}, it 
is clear that the analysis in the former is more reliable, with a larger number of data points used and an overall much weaker size dependence. The above 
result from the hydrodynamic relationship is still likely somewhat affected by systematical errors, as the behavior still appears to be flattening out 
more than the fit suggests, and the fit even including the 6th-order term is only marginally good, with $\chi^2/{\rm dof} \approx 1.7$. It would clearly 
be desirable to go to larger system sizes, but given the much better behavior of the winding number data it is already clear that this is the 
preferred method. 

\section{Bilayer Heisenberg model}
\label{sec:bilayer}

We now consider the $S=1/2$ bilayer Heisenberg model which is a prototypical system to study quantum phase transitions in 2D.\cite{singh88,millis93,sandvik94} 
The Hamiltonian is given by
\begin{equation}
H = J_1 \sum_{\langle i,j\rangle } ({\bf S}_{i,1} \cdot {\bf S}_{j,1}+{\bf S}_{i,2} \cdot {\bf S}_{j,2})
+ J_2 \sum_{i=1}^{L^2} {\bf S}_{i,1} \cdot {\bf S}_{i,2},
\label{hbilayer}
\end{equation}
where $\mathbf{S}_{i,a}$ represents a $S=1/2$ spin operator at site $i$ of layer $a$ ($a=1,2$), and $\langle i,j \rangle$ denotes a pair of nearest neighbor 
sites on the square lattice of $L \times L$ sites with periodic boundary conditions. Both the couplings $J_1,J_2$ are positive (antiferromagnetic). As the 
ratio $g=J_2/J_1$ is increased, there is a destruction of long-range N\'eel order at a critical $g_c$, beyond which the system enters a disordered state 
with no broken symmetries. The best value available for the location of the critical point is $g_c=2.5220(1)$ \cite{wang06} and we will use this
value below.

\subsection{Velocity from winding numbers}

\begin{figure}
\center{\includegraphics[width=8cm, clip]{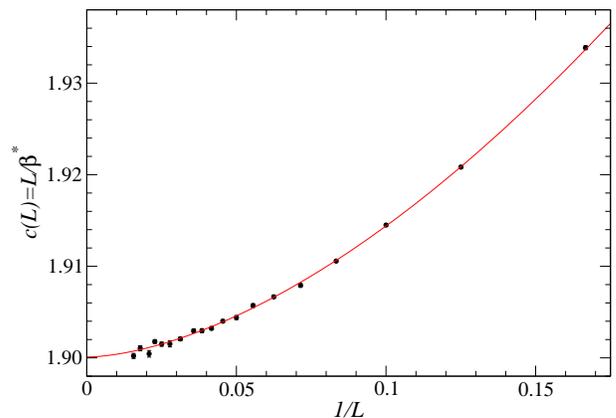}}
\vskip-2mm
\caption{(Color online) Size dependence of the velocity computed using the winding-number cubic criterion for the bilayer Heisenberg model 
at  the estimated critical coupling ratio $g=2.5220$. The curve shows a quadratic fit of the form $c(L)=c+b/L^a$ to the $L \ge 8$ data,
which gives the extrapolated velocity $c=1.9001(2)$.}
\label{cwind_bi}
\vskip-4mm
\end{figure}

Results for the velocity based on the cubic winding-number criterion at the critical point of the bilayer
are shown in Fig.~\ref{cwind_bi}. Here we find that a non-integer power-law correction describes the data very well for system sizes $L\ge 6$. A fit of 
the form $c(L)=c+b/L^a$ to the $L \ge 8$ data gives $c=1.9001(2)$. Note that the fitted curve also goes through the $L=6$ data point even though this 
point was not included in the fit. This adds to our confidence of this power-law correction. The value of $c$ is a few percent smaller than the spin-wave
result \cite{sandvik94} $c_{\rm sw} =1.96$ including $1/S$ corrections at $g=2.51$ and $c_{\rm sw}\approx 2.03$ obtained \cite{morr95} to order$1/S^2$.
A more sophisticated  treatment beyond conventional spin-wave theory, including the effects of longitudinal fluctuations close to the critical point, gives
$c = 1.78$ when the expression $c=0.705 \times g$ below Eq.~(10) of Ref.~\onlinecite{morr95} is evaluated with $g_c=2.522$. While these analytical 
values may appear reasonably close to (deviating by a few percent from) the presumably correct numerical value we have obtained here, the deviations 
are still much larger than in the case of the ordered 2D Heisenberg model, where the error of spin-wave result is only $0.3\%$, as discussed 
in Sec.~\ref{sec:winding2d}.

The exponent of the correction in the fit in Fig.~\ref{cwind_bi} is $a=1.67(4)$. It is not clear to us how this exponent relates to the standard critical exponents 
of the 3D O(3) universality class of the transition, but the relatively large value (larger than $1/\nu\approx 1.41$) suggests that it may involve subleading 
exponents.

\subsection{Velocity from gaps}

Since the bilayer has two sites per unit cell ($a=1,2$ for each square-lattice point $i$), the triplet excitations have an extra label $k_z$ in $\mathbf{k}$ 
space which denote the in-phase ($k_z=0$) and out-of-phase ($k_z=\pi$) spin excitations of the two layers, respectively. In the magnetically ordered N\'eel 
phase, the triplet excitations are gapless at $(0,0,0)$ and $(\pi,\pi,\pi)$, with the lowest excitation being at $(\pi,\pi,\pi)$ for a finite system. The 
spectrum is linear in the neighborhood of both the gapless points, which defines the corresponding spin wave velocity $c$. For a continuous phase transition, 
the spin wave velocity $c$ scales as~\cite{chubukov94}
\be
c \propto (g-g_c)^{\nu (z-1)}, 
\ee  
where $z$ is the dynamical exponent and $\nu$ is the correlation length exponent. Thus, for a $z=1$ transition as is the case for the Heisenberg bilayer at $g=g_c$, 
the velocity is regular at the critical point. 

\subsubsection{Critical point} 

\begin{figure}
\center{\includegraphics[width=8cm, clip]{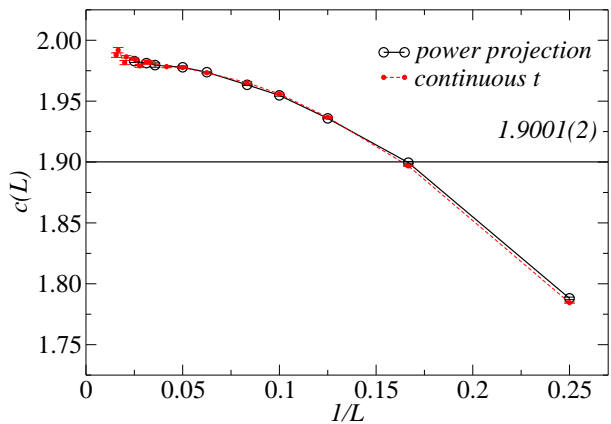}}
\vskip-2mm
\caption{(Color online) The velocity estimator (\ref{clbil}) for the Heisenberg bilayer at its critical point shown as a function of $1/L$. Results based
on both power-projection and exponential projection (continuous time) are shown and agree within error bars. The velocity estimate obtained from the winding numbers 
(Fig.~\ref{cwind_bi}) is shown as the horizontal line.}
\label{cgaps_bi}
\vskip-2mm
\end{figure}

For measuring the velocity of the critical modes we first study the triplet gap at $(\pi+2\pi/L,\pi,\pi)$. From the linearly of the spectrum at the 
critical point, we define the velocity estimator in analogy with the single-layer case (\ref{cl2d}) as 
\be
c(L)=\Delta (\pi+2\pi/L,\pi,\pi)L/(2\pi).
\label{clbil}
\ee
The behavior of this quantity as a function of $L$ is shown in Fig.~\ref{cgaps_bi}. Unlike the case of the Heisenberg chain (Fig.~\ref{velexact1d}) and the single-layer
(Fig.~\ref{neelgaps}), the velocity estimate for the Heisenberg bilayer at criticality is notably higher than winding-numbers estimate, by about $5 \%$ as 
$L \rightarrow \infty$. To extract the gap needed in Eq.~(\ref{clbil}) we have used both the power-projection QMC method with $\delta$-function fits to the correlation 
function, as described in Sec.~\ref{sec:ct}, and the generalized moment method applied to exponential-projector QMC data, as detailed in Sec.~\ref{sec:moments}. 
As seen in in Fig.~\ref{cgaps_bi}, the two methods give results that agree fully within statistical error.

At first sight, potential reason for the disagreement with the winding-number result could be an over-estimation of the gap extracted from the correlation 
function at the critical point.  Consider the full dynamic spin structure factor defined as
\be
S_\mathbf{k}(\omega) = \sum_m |\langle m|S^z(\mathbf{k})|0 \rangle|^2 \delta(\omega+E_0-E_m). 
\ee
In the N\'eel phase the structure factor has a $\delta$-function of weight $A_d(\mathbf{k})$ at an energy $\omega(\mathbf{k})$ which represents the magnon mode, as
well as a continuum $A_c(\mathbf{k},\omega)$ which does not extend below $\omega(\mathbf{k})$ and which decays rapidly to zero as $\omega \rightarrow \infty$:
\be
S_\mathbf{k}(\omega) = A_d(\mathbf{k})\delta \bigl (\omega-\omega(\mathbf{k}) \bigr)+ A_c(\mathbf{k},\omega). 
\ee
The velocity derived from $\omega(\mathbf{k})$ in the vicinity of $(\pi,\pi,\pi)$ is a regular function of $g-g_c$ and is by definition the correct velocity at the 
critical point. However, the weight in the $\delta$-function $A_d(\mathbf{k})$ (that represents the magnon mode) also smoothly goes to zero as the critical point 
is approached, with the spectrum evolving into purely a continuum reflecting the overdamped critical magnons. In our fitting scheme [see Eqs.~(\ref{gaps1}) and 
(\ref{gaps2})] used with the power-projection QMC method, where the spectral function is represented by a small number of $\delta$ functions, $\omega(\mathbf{k})$ 
may be {\it over-estimated}, especially for large system sizes, when $A_d$ is very small close to the critical point. The velocity would then also be over-estimated,
as it is in Fig.~\ref{cgaps_bi} (assuming that the winding-number result is correct). A similar distorting effect of the continuum may be expected also with the 
generalized moment method used with the exponential-projection QMC data. However, the $\delta$-function fits are stable with respect to the number of $\delta$-functions 
used, and the extrapolations used with the generalized moment method are also stable. We do not see any evidence of remaining effects that could account for
an overestimation as large as in Fig.~\ref{cgaps_bi}. The perfect match between the two methods within their statistical errors also gives us confidence that 
the gaps are determined correctly and the reason for the disagreement with the winding-number method must be sought elsewhere.

\begin{figure}
\center{\includegraphics[width=8cm, clip]{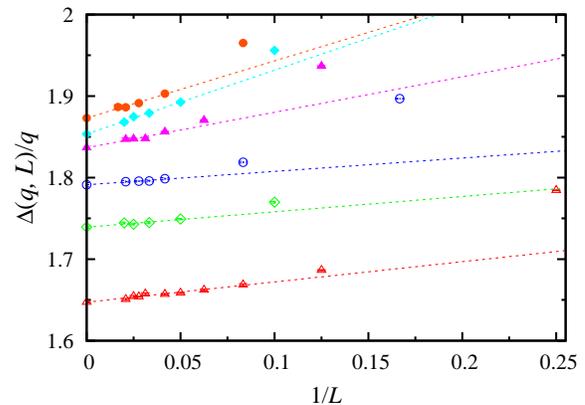}}
\vskip-2mm
\caption{(Color online) Momentum dependent gap, with $\mathbf{q}$ defined relative to $(\pi,\pi,\pi)$, divided by the respective $q$.
The data are shown versus $1/L$ for $q=2\pi/4$ (open triangles), $2\pi/5$ (open diamonds), $2\pi/6$ (open circles), $2\pi/8$ (solid triangles), $2\pi/10$ (solid diamonds), 
$2\pi/12$ (solid circles). Linear fits in $1/L$ are shown for each momentum, with only data for sufficiently large $L$ included for each $q$.}
\label{fig:c-bl-1}
\vskip-2mm
\end{figure}

\begin{figure}
\center{\includegraphics[width=8cm, clip]{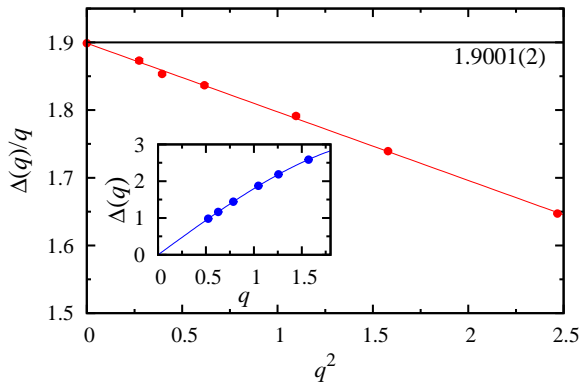}}
\vskip-4mm
\caption{(Color online) Extrapolation of the critical bilayer velocity using the gaps for $q \rightarrow 0$. The infinite-size velocity is estimated as the
limiting value $c=1.899(2)$, which is consistent with the estimation from the winding numbers, $c=1.9001(2)$, indicated by the horizontal line. A linear 
function in $1/q^2$ is used for the extrapolation. The inset shows the calculated dispersion.}
\label{fig:c-bl-2}
\vskip-2mm
\end{figure}

To search for a possible flaw in the velocity estimation based on Eq.~(\ref{clbil}), we next analyze the details of the dispersion relation around $(\pi, \pi, \pi)$, 
using gaps extracted with the generalized moment method. The dispersion should be an asymptotically linear function of the wave number, $\Delta(\mathbf{q}) = c q$, where 
$\mathbf{q}$ is the momentum relative to the gapless point $(\pi, \pi, \pi)$;
\begin{equation}
\mathbf{q} = \mathbf{k}-(\pi,\pi,\pi).
\end{equation}
We should then have $\lim_{q\rightarrow 0} \lim_{L \rightarrow \infty} \Delta(\mathbf{q},L) / q=c$. There could potentially be an issue with the order of the limits 
$q \to 0$ and $L \to \infty$, which with Eq.~(\ref{clbil}) are taken simultaneously as we use $q=2\pi/L$.

To investigate the formally correct limit of taking $L \to \infty$ first and then $q \to 0$, we analyze the gaps at fixed $q$ and varying $L$, to find the 
corresponding gap values as a function of the momentum in the thermodynamic limit. The $L$ dependence of the gaps at several different $q$-values are shown 
in Fig.~\ref{fig:c-bl-1}. We observe that the size correction to the gap for sufficiently large $L$ is linear in $1/L$ and therefore extrapolate the data to 
infinite size using simple line fits. Using these gaps $\Delta(q)$, we finally extrapolate the velocity as $c=\lim_{k\rightarrow0} \Delta(q)/q$, as shown 
in Fig~\ref{fig:c-bl-2}. The behavior is fully consistent with a linear dispersion in the limit $q \to 0$, and empirically we find that the leading correction 
is of order $q^2$ (i.e., the correction to the gap $\Delta \sim q$ is of order $q^3$). The final result for the velocity extracted this way 
is $c=1.899(2)$, which is fully consistent with the result based on the winding numbers. Thus, the reason for the previous disagreement is indeed that
Eq.~(\ref{clbil}) does not represent the correct order of limits, and there are no flaws in the extraction of the gaps.

This more detailed analysis also shows the reason for the over-estimation of the velocity based on Eq.~(\ref{clbil}). As is clear from Fig~\ref{fig:c-bl-1}, 
the gaps converge to their infinite size values as $1/L$ for sufficiently large $L$. We observe that the gaps at fixed momentum $k$, where $k$ is 
close to but not equal to $(\pi,\pi,\pi)$, follows the following behaviour for large $L$:
\begin{eqnarray}
\Delta(q,L) &=& \Delta(q,L \rightarrow \infty)+ \frac{A(q)}{L^2} \mbox{~~~}\mathrm{(gapless)} \nonumber \\
\Delta(q,L) &=& \Delta(q,L \rightarrow \infty)+ \frac{B(q)}{L} \mbox{~~~}\mathrm{(critical)} \nonumber \\
\Delta(q,L) &=& \Delta(q,L \rightarrow \infty)+ a(q){\rm e}^{-b(q)L} \mbox{~~~}\mathrm{(gapped)} \nonumber
\end{eqnarray}
In the gapless and the gapped phases, the size correction decays sufficiently rapidly, so that $\Delta(q,L \rightarrow \infty) = ck$. However, at the 
critical point, the estimator $\Delta(q,L)L/2\pi$ instead converges to $c+B(q \rightarrow 0)/2\pi$. 

\subsubsection{Paramagnetic phase} 

\begin{figure}
\center{\includegraphics[width=8.75cm, clip]{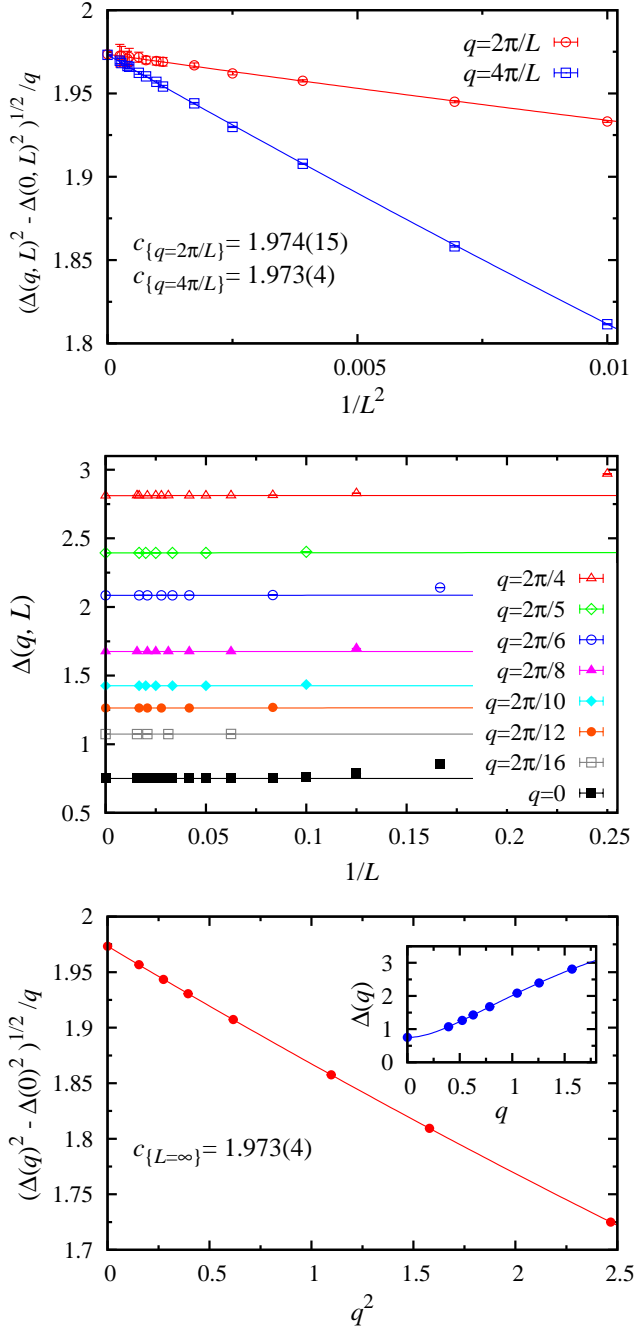}}
\vskip-4mm
\caption{(Color online) Velocity extraction for the bilayer at $g=3$ (paramagnetic phase). The top panel shows the size
convergence of velocity estimates based on two momenta close to the antiferromagnetic momentum ($q$ being the deviation from this momentum). The middle panel 
shows the system-size dependence of the gaps at several values of $q$. The bottom panel shows the infinite-size extrapolated velocity defined versus $q$, obtained from 
the fits in the middle panel and graphed versus $q^2$. The  $q \to 0$ extrapolation by a polynomial (red curve) agrees with the different order of the
limits taken in the top panel. The inset shows the dispersion relation.}
\label{gapbilayerg3}
\end{figure}

For completeness we next extract the velocity of the propagating triplet excitations, or triplons, also in the quantum-disordered phase of the bilayer. 
We choose two points well away from $g=g_c$, at $g=3$ and $g=4$. The lowest triplet gap at $(\pi,\pi,\pi)$ now converges to a finite ($g$ dependent) 
value as $L \rightarrow \infty$, because the paramagnetic phase is gapped. In the vicinity of $(\pi,\pi,\pi)$, the spectrum is expected to 
behave generically as 
\be
\Delta(\mathbf{q}) = \sqrt{\Delta_0^2+c^2q^2} \approx \Delta_0+\frac{c^2 q^2}{2 \Delta_0}, 
\label{gappeddis}
\ee
where $\Delta_0$ denotes the triplet gap at wavevector $(\pi,\pi,\pi)$, $c$ is the velocity of the gapped triplons, and $q$ is again measured relative 
to $(\pi,\pi,\pi)$. We use both the approaches (ways of taking $q\to 0$ and $L \to \infty$) discussed above to estimate the triplon velocity $c$ and they 
give consistent results in the gapped phase; see Fig~\ref{gapbilayerg3} for the analysis at $g=3$.  We obtain $c=1.973(4)$ and $c=2.159(6)$ 
for $g=3$ and $g=4$, respectively 

The simple estimator (\ref{clbil}) should give the correct velocity in the gapped phase, as discussed above, and we further check for consistency of the 
approach by also analyzing the gap at $q=4\pi/L$, in addition to the smallest momentum $q=2\pi/L$ (both extrapolations shown in Fig~\ref{gapbilayerg3}, upper panel).
The same velocity is also obtained using these extrapolated gap values representing the limit $L \to \infty$ before $q\to 0$, as shown in Fig~\ref{gapbilayerg3}, 
bottom panel.    

\begin{figure}
\center{\includegraphics[width=8cm, clip]{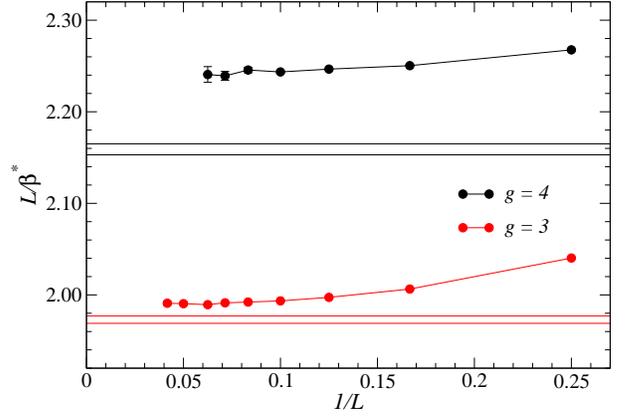}}
\vskip-1mm
\caption{(Color online) The behavior of the inverse temperature $\beta^*$ at which the cubic criterion is satisfied for the bilayer at
two couplings in the paramagnetic regime. The method does not produce the correct velocity when $g>g_c$. The values plus-minus one error bars obtained using
the gap method is shown with the double horizontal lines.}
\label{windg3g4}
\vskip-1mm
\end{figure}

For completeness, we also show the velocity estimate $c(L)$ obtained from the winding numbers with the cubic criterion in Fig.~\ref{windg3g4}. $c(L)$ converges (exponentially fast for large $L$) to a finite value as $L \rightarrow \infty$. However, this estimate gives an incorrect (higher) velocity in the paramagnetic phase. This is expected as the triplons are not linearly dispersing excitations [see Eq.~(\ref{gappeddis})]. The error in the estimate increases with the distance of $g$ from the critical point, which is because $c$ is a regular function of $(g-g_c)$ from both sides and the winding number estimator does work at the critical point.      

\section{Quantum-critical scaling at $T>0$}
\label{sec:bilayerT>0}

With $c$ of the critical bilayer determined to high precision, we have an opportunity to test in detail quantum-critical scaling predictions 
based on large-$N$ calculations for the O($N$) model, which for $N=3$ should describe the universal critical behavior.\cite{chakravarty89,chubukov94} 
The $T=0$ quantum-critical point influences the behavior of the system in a wide ``fan'' in the ($g,T$) plane, with cross-overs to different low-$T$ 
behaviors away from $g_c$ set by the spin stiffness (for $g<g_c$) and the spin gap (for $g>g_c$). Here we will test 
results for the temperature dependence of the uniform magnetic susceptibility $\chi(T)$ (per unit cell),
\begin{equation}
\chi = \frac{1}{L^2}\left \langle \left (\sum_{j=1}^{2}\sum_{i=1}^{L^2} S^z_{i,j} \right )^2 \right \rangle,
\label{chidef}
\end{equation}
and the specific heat,
\begin{equation}
C = \frac{d}{dt}E(T),
\label{chidef2}
\end{equation}
where the internal energy $E=\langle H\rangle/L^2$ per unit cell is computed with the bilayer Hamiltonian (\ref{hbilayer}). In the SSE
method each configuration has a fixed $z$ magnetization (since this is a conserved quantity) and $\chi$ is evaluated directly according
to Eq.~(\ref{chidef}). $C$ can be computed using an exact estimator based on the fluctuations of the number of operators in the sampled 
operator strings. In practice, however, this estimator is very noisy at low temperatures and we will therefore instead analyze the energy,
which is simply related to the average number of SSE operators.

\subsection{Results of large-N calculations}

The large-$N$ approximations give the following leading-order low-$T$ forms of the susceptibility and the specific heat at the critical coupling:\cite{chubukov94}
\begin{equation}
\chi'(T) = \frac{\lambda}{\pi c^2}T,
\label{chitp}
\end{equation}
where the constant $\lambda=1.0760$, and
\begin{equation}
C'(T) = \frac{36 \zeta(3)}{5\pi c^2}T^2,
\label{ctp}
\end{equation}
where $\zeta(3)=1.20206$. Since we in practice compute and analyze the internal energy, we will compare with the $T$-integrated
version of (\ref{ctp});
\begin{equation}
E'(T) = E_0+\frac{12 \zeta(3)}{5\pi c^2}T^3.
\label{etp}
\end{equation}
We use primes on the symbols above to indicate that these are not expected to be exact forms. Several tests have been
reported in the literature for different variants of dimerized Heisenberg models and the leading power laws above have been
confirmed. However, quantitatively the degree of agreement has not been that well established, because $c$ has also typically been
extracted from quantities relying on the large-$N$ expansions, instead of using an independently determined unbiased value. 

One can also consider the Wilson ratio,
\begin{equation}
W = \frac{\chi T}{C},
\end{equation}
which has the advantage that its approximant based on Eqs.~(\ref{chitp}) and (\ref{ctp}) does not involve the velocity. From the
above expressions for $\chi$ and $C$ one obtains
\begin{equation}
W' = \frac{5\lambda}{12 \zeta(3)} \approx 0.1243.
\label{wprime}
\end{equation}
A value differing from this prediction by only about $2\%$ was recently reported in Ref.~\onlinecite{sandvik11} based on large-scale studies of 
a single-plane model with columnar dimerization; $W=0.1262(6)$. Here our main aim is to obtain precise estimates for the prefactors $a$ and $b$ of 
the leading low-$T$ forms $\chi = aT$, $C=bT^2$ and compare these with the predicted values in with the value of the velocity $c=1.9001(2)$, 
as determined in the previous section.

\subsection{Susceptibility}

\begin{figure}
\center{\includegraphics[width=8.25cm, clip]{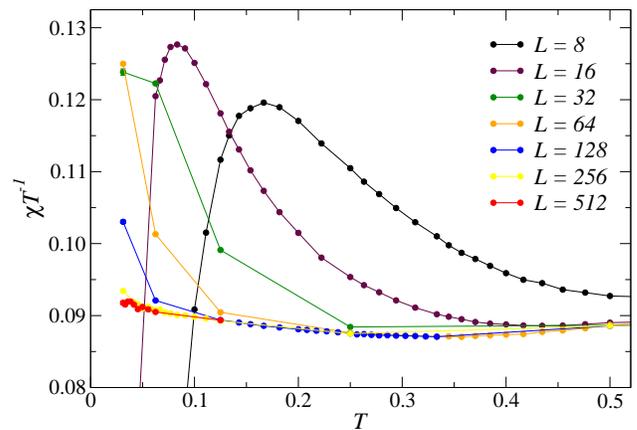}}
\vskip-2mm
\caption{(Color online) Susceptibility per unit cell divided by the temperature of the bilayer at $g=2.5220$.
Result for several different system sizes are shown to illustrate the finite size effects. For any finite $L$ the susceptibility vanishes 
when $T \to 0$, at a temperature scale set by the finite size gap $\Delta$, which at criticality scales as $\Delta \sim 1/L$. The lines
connecting points are guides to the eye.}
\label{susc_l}
\vskip-1mm
\end{figure}

\begin{figure}
\center{\includegraphics[width=8cm, clip]{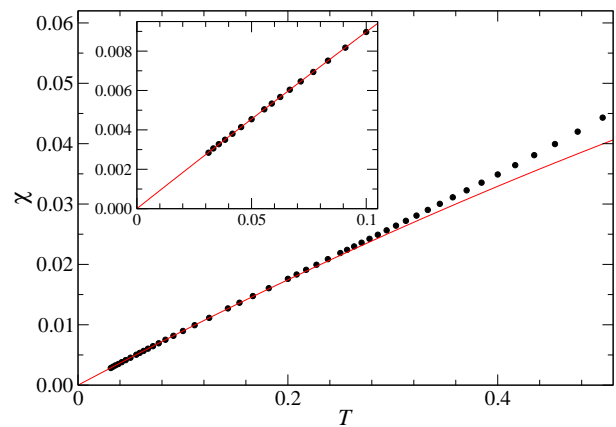}}
\vskip-2mm
\caption{(Color online) The susceptibility per spin of the bilayer Heisenberg model at coupling ratio $s=2.5221$. The
points are SSE results (with error bars much smaller than the symbols) and the curves are second-order polynomial fits
to the data for $T/J_1 \le 0.1$.}
\label{xfit1}
\vskip-1mm
\end{figure}

We need the susceptibility per unit cell in the thermodynamic limit. The finite-size effects are illustrated in Fig.~\ref{susc_l} for
a coupling ratio close to the quantum-critical value. To make the graph clearer we have graphed $\chi /T$, which also is appropriate 
considering that we are here aiming to extract the size of the $T$-linear term in $\chi$. Since the magnetization is conserved and there 
is a finite size singlet-triplet gap $\Delta \sim 1/L$, the susceptibility for a given finite $L$ vanishes exponentially at a temperature 
scale $T \sim 1/L$. Interestingly $\chi/T$ exhibits a prominent peak preceding the eventual drop to zero. The results indicate that
we can compute the susceptibility without noticeable (within statistical errors) finite-size effects down to $T/J_1 = 1/32$ by using $L=512$ 
lattices at the lowest temperatures (while smaller systems can be used at higher temperatures). 

Below we show results for a range of temperatures representing the thermodynamic limit. In addition to calculations at $g=2.5220$, we have also 
considered couplings one standard deviation of the $g_c$ estimate away from this point, i.e., $g=2.5219$ and $2.5221$. Based on these calculations
we observe that the critical point should be very close to $2.5221$. In Fig.~\ref{xfit1} we analyze the $T$ dependence of the susceptibility.
At $g=2.5221$ a second-order polynomial fit works well with data for $T/J_1 \le 0.1$ ($\chi^2/{\rm dof} \approx 1.2$ with 16 data points) 
and gives zero intercept within the statistical error of the fit; $0.000004(16)$. The slope $a$ in the expected leading-order form
$\chi = aT$ is $a=0.0922(3)$. If the intercept is assumed to be $0$ and the form $\chi = aT + bT^2$ is used, the resulting
slope is $0.09224(9)$. A second-order fit to data at at $g=2.5220$ and the same range of temperatures (not shown in the figure) gives a positive 
intercept about two error bars away from $0$; $0.000032(14)$. The slope is $a=0.0916(6)$, not much different from the result for $g=2.5221$. 
Although the differences between these data sets are small, they, along with results for $g=2.5219$, suggest that the critical point is
closer to $g=2.5221$ than to $g=2.5220$, which is still in agreement within error bars with the result $g-=2.5220(1)$ stated in 
Ref.~\onlinecite{sandvik11}. Based on the present susceptibility results we estimate $g=2.52210(5)$.

\begin{figure}
\center{\includegraphics[width=8cm, clip]{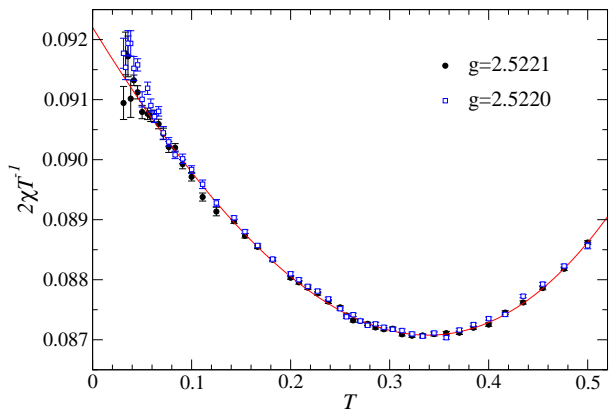}}
\vskip-2mm
\caption{(Color online) Twice the susceptibility divided by the temperature of the bilayer Heisenberg model at two coupling ratios close to 
the quantum-critical point. The solid curve is a third-order polynomial fit to the $g=2.5221$ (which we estimate is closer to $g_c$) data.} 
\label{xfit2}
\vskip-2mm
\end{figure}

Fig.~\ref{xfit2} shows the susceptibility per unit cell with the temperature divided out at $g=2.5220$ and $2.5221$, along with 
a fit to the data at the latter coupling. A third-order polynomial describe all the data well for all temperatures up to about $T/J_1=0.5$.
The intercept is completely consistent with the results for the slope obtained in Fig.~\ref{xfit1}: $a=0.09220(5)$. Given that this fit 
includes the largest amount of data we take the result as our final estimate of the susceptibility prefactor. We can now compare it with
the predicted value from Eq.~(\ref{chitp}), which with the value of $c$ extracted in Sec.~\ref{sec:bilayer} is $a'=0.09487(2)$, i.e., $2.9\%$ higher 
than our estimate. This close agreement is quite remarkable, considering that $a'$ is based on a leading $1/N$ calculation evaluated at $N=3$.

\subsection{Specific heat}

\begin{figure}
\center{\includegraphics[width=8cm, clip]{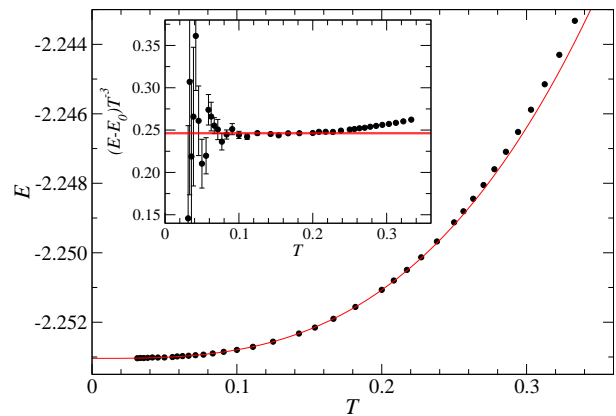}}
\vskip-2mm
\caption{(Color online) The internal energy of the bilayer Heisenberg model at coupling ratio $g=5.5221$ (within the error bars of the estimated $g_c$), 
along with a fit of the form $E=E_0+aT^3$ to the $T \le 0.2$ data. The inset shows the data after subtraction of the ground-state energy
and dividing by $T^3$. The horizontal line shows the prefactor of the cubic correction (with the line thickness corresponding approximately 
to $\pm$ one error bar) from the fit in the main graph.}
\label{efit2}
\vskip-2mm
\end{figure}

The internal energy is graphed in Fig.~\ref{efit2}. The form $E=E_0+bT^3$ describes well the data for $T \le 2$ and a fit gives $E=-2.253040(1)$ 
and $b=0.2462(6)$. The predicted factor  from Eq.~(\ref{etp}) is $b'=0.25435(5)$, which is $3.3\%$ higher than our estimate. 

\subsection{Wilson ratio}

The Wilson ratio with the exact prefactors $a$ and $b$ in the asymptotic forms $\chi = aT$ and $E=E_0+bT^3$ ($C=3bT^2$) is $W=3a/b$. With the 
values of $a$ and $b$ determined above we obtain $W=0.1248(3)$. This value agrees reasonably well with the less accurate (with larger error bar) value 
obtained in Ref.~\onlinecite{sandvik11}, being smaller by about two error bars. It is in remarkably good agreement with the $1/N$ value in Eq.~(\ref{wprime}), 
deviating by only one error bar. In other word, at the $95\%$ confidence level (about two error bars), the $1/N$ estimate agrees with the true value to within 
about $0.5\%$ or better.

\section{Summary and discussion}
\label{sec:discussion}

We have here presented several methods to extract the velocity of excitations in quantum spin models simulated using QMC techniques. The methods are 
fundamentally not new, but we presented several technical solutions to improve their practical utility, and carried out extensive tests with the goal 
of testing their precision in practice.

The method of computing the lowest 
excitation energy for given momentum using the long-time behavior of imaginary-time correlations directly probes the dispersion relation and is, thus, 
directly connected to the definition of the velocity. This is also the most complicated method in practice, as it requires significant efforts to obtain
high-precision results for appropriate correlation functions and to analyze them, trying to avoid contamination by excited states in the gap estimation. 
We here employed two different methods to compute the correlations, using ground-state projector and finite-temperature QMC simulations. We also presented 
two different ways to extract the gap using fitting procedures sensitive to the slowest decaying component of the correlation functions. The good agreement 
of the velocity obtained in this way with the other approaches in the limit of large system sizes show that the gap estimators are unbiased (and we also
demonstrated this directly for small system sizes where exact results can be obtained). We expect these gap-extraction  procedures to be useful also for 
many other systems and studies.

The method of using winding number fluctuations to define a cubic space-time geometry fundamentally relies on the low-energy physics of systems
with dynamic exponent $z=1$. The method requires calculations for several temperatures in the neighborhood of the target temperature $T^*$ at 
which the temporal and spatial winding number fluctuations are equal, or, in principle, one could use a re-weighting technique and only do a single
simulation very close to the estimated crossing point, though we expect that it should be statistically advantageous to do several independent runs
as more independent data are collected and the statistical analysis (curve fitting and error estimate) is much simpler. Though doing several simulations 
may seem like a drawback, in practice the procedure for extracting the crossing temperature are very straight-forward and based on our work presented 
here this is the preferred method to extract the velocity. As noted in a previous application of this method,\cite{jiang11a} the convergence with 
the system size is very rapid, but here we still found it necessary to use a final extrapolation using a power-law correction (with integer or non-trivial
leading power in a long-range ordered and critical system, respectively), to avoid any remaining size effects.

We also presented a modification of the standard hydrodynamic relationship between the velocity, spin stiffness, and susceptibility, computing
the latter at non-zero momentum $q=2\pi/L$ in order to obtain a useful finite-size estimate of the velocity (with the standard relationship at
$q=0$ diverging when $T \to 0$ in a finite system in QMC simulations). This method also exhibits good convergence properties, though in practice
it is still not as powerful as the winding-number method.

Finally, we extracted a high-precision estimate of the velocity in a critical bilayer model and used it in combination with QMC calculated thermodynamic
properties to test field-theory predictions based on $1/N$ expansions of the $2+1$ dimensional O$(3)$ model. The independently extracted velocity allowed
us to precisely determine the deviations of the predicted overall factors in the temperature dependence of the susceptibility and the specific heat; the
$1/N$ results (with $N=3$) deviate from the true values by about 3\%. The Wilson ratio, where the velocity dependence cancels out, agrees with the
$1/N$ value to within our $0.3\%$ error bar (and we cannot establish the level of agreement beyond the level of the error bar).

\begin{acknowledgments}
AS acknowledges the computational resources of the Max Planck Institute for the Physics of Complex Systems, Dresden, 
where some of the simulations were performed. HS acknowledges the computational resources of the Institute for
Information Management and Communication, Kyoto University, and Computing and Communications Center, Kyushu University, 
through the HPCI System Research Project (No.\,hp140204). The work of HS was supported by CMSI in SPIRE from MEXT, 
Japan, and the JSPS Postdoctoral Fellowships for Research Abroad. HS also acknowledges the use of the ALPS library 
\cite{ALPS2011s} for simulation scheduling. The work of AWS was supported by the NSF under Grant No.~DMR-1410126 and 
by the Simons Foundation. 
\end{acknowledgments}

\appendix

\section{Nature of the systematical errors in the $(n,\beta_{\rm int})$ gap estimator}

We will show analytically the systematic error of the gap
estimator~(\ref{eqn:gap_Fourier}) and the
limits~(\ref{eqn:gap_Fourier_limits}) in more details, which was shown in Sec.~\ref{sec:ct}.  Let us recall the dynamical correlation function in the exponential-form projection:
\begin{align}
C(\tau, \tau_0) &= \frac{1}{Z'} \langle \psi_t | e^{ - ( \beta - \tau - \tau_0 ) H} A^{\dagger} e^{-\tau H} A e^{- \tau_0 H} | \psi_t \rangle \nonumber \\
&= \frac{1}{Z'} \sum_{\ell, p, q} b_{\ell, p, q} \, e^{-\beta E_p} e^{- \tau ( E_{\ell} - E_p )} e^{ - \tau_0( E_q - E_p )} \nonumber ,
\end{align}
where $Z' = \langle \psi_t | e^{-\beta H} | \psi_t \rangle $, $| \psi_t \rangle = \sum_q c_q | q \rangle$, $H | q \rangle = E_q | q \rangle $, and $b_{\ell,p,q}= \bar{c}_p c_q \langle p | A^{\dagger} | \ell \rangle \langle \ell | A | q \rangle $.
The Fourier transformed correlation function at frequency $\omega_k= 2 \pi k / \beta_{\rm int}$ $(k \in \mathbf{Z})$, is expressed as 
\begin{align}
\tilde{C}{(\omega_k)} &= \int_0^{\beta_{\rm int}} d\tau \, C(\tau, \tau_0) e^{i \tau \omega_k} \\
&= \left\{
\begin{array}{l}
\displaystyle \frac{1}{Z'} \sum_{\ell, p} g_{\ell, p} \frac{ \Delta_{\ell, p} + i \omega_k }{ \Delta_{\ell, p}^2 + \omega_k^2 } \quad ( \omega_k \neq 0 ) \\
\displaystyle \frac{1}{Z'} \{ \sum_{E_{\ell} \neq E_p} \frac{ g_{\ell,p} }{ \Delta_{\ell, p} } + \sum_{E_{\ell} = E_p} d_{\ell,p} \beta_{\rm int} \} \quad ( \omega_k = 0 ), \nonumber
\end{array}
\right.
\end{align}
where $\beta_{\rm int} \leq \beta - \tau_0$, $g_{\ell, p} = d_{\ell, p} ( 1 - e^{ - \beta_{\rm int}
  \Delta_{\ell, p}} ) $, $d_{\ell, p}= e^{- ( \beta - \tau_0 ) E_p}
\sum_q b_{\ell, p, q} e^{- \tau_0 E_q }$, $\Delta_{\ell, p}=E_{\ell} -
E_p$. Note that the imaginary part for $\omega_k \neq 0 $ does not
vanish as finite-temperature case because the correlation function
$C(\tau,\tau_0)$ is not periodic.

The gap estimator~(\ref{eqn:gap_Fourier}) is rewritten as
\begin{align}
\frac{ \hat{\Delta}_{(n, \beta_{\rm int})} }{ \Delta_1} &= \frac{  1 + R_n (\beta_{\rm int}) }{ 1 + \displaystyle  F_n (\beta_{\rm int})  + D_n (\beta_{\rm int}) }  \label{eqn:gap-n-beta} \\
  \rightarrow  \ 1 + & \sum_{\ell > 1} \left[ \frac{ b_{\ell} }{b_1} \left( \frac{\Delta_1}{\Delta_{\ell}} \right)^{2n} \! \! \! \! \! \! + O\left( \left( \frac{\Delta_1}{\Delta_{\ell}} \right)^{2n+1} \right) \right] ( \beta_{\rm int} \rightarrow \infty ) \nonumber ,
\end{align}
where
\begin{align}
R_n (\beta_{\rm int}) &= \sum_{ \{ \ell, p \} \in S } \left( \frac{g_{\ell, p}}{g_{1,0}} \right) \frac{ h(n, \beta_{\rm int}, \ell, p ) }{ h( n, \beta_{\rm int}, 1, 0 ) } \nonumber \\
F_n (\beta_{\rm int}) &= \sum_{ \{ \ell, p \} \in S } \left( \frac{g_{\ell, p}}{g_{1,0}} \right) \frac{ h(n, \beta_{\rm int}, \ell, p ) }{ h( n, \beta_{\rm int}, 1, 0 ) } \left( \frac{ \Delta_1 }{ \Delta_{\ell, p} } \right) \nonumber \\
D_n (\beta_{\rm int}) &=  \sum_{E_{\ell} = E_{p} }  x_{n,0,0} \left( \frac{d_{\ell, p}}{g_{1,0}} \right) \beta_{\rm int} \Delta_1 \frac{1}{ h(n, \beta_{\rm int}, 1, 0) } \nonumber
\end{align}
\begin{align}
h(n, \beta_{\rm int}, \ell, p) = (-1)^n \omega_1^{2n} \prod_{k=1}^n \frac{ 1 }{ \Delta_{\ell, p}^2 + \omega_k^2 }  \nonumber ,
\end{align}
$b_{\ell} \equiv
b_{\ell,0,0}= | c_0 \langle \ell | A | 0 \rangle |^2$, and
$\Delta_{\ell} = E_{\ell} - E_0 $.
For the summation of $R_n$ and $F_n$, the set $S$ is such that $E_{\ell} \neq E_{p}$ except $(\ell, p) = ( 1, 0 )$. Then
\begin{align}
\lim_{n \rightarrow \infty} \lim_{\beta_{\rm int}, \tau_0 \rightarrow \infty} \hat{\Delta}_{( n,\beta_{\rm int} )} = \Delta_1 .
\end{align}
Let us consider taking the limit $n \rightarrow \infty$ before
$\beta_{\rm int}, \tau_0 \rightarrow \infty$.  Using the product
expansion form of the hyperbolic function, $\sinh ( \pi z ) = \pi z
\prod_{k=1}^{\infty} ( 1 + z^2 / k^2 )$, we obtain
the limiting form as
\begin{align}
\frac{ \hat{\Delta}_{(n, \beta_{\rm int})} }{ \Delta_1} = \frac{  1 + R_{\infty} (\beta_{\rm int}) }{ 1 + \displaystyle  G_{\infty} (\beta_{\rm int}) }  + O \left( \frac{1}{n} \right) \label{eqn:n-inf} ,
\end{align}
where
\begin{align}
R_{\infty} (\beta_{\rm int}) &= \sum_{ \{ \ell, p \} \in S }
\left( \frac{d_{\ell, p}}{d_1} \right) \left( \frac{ \Delta_{\ell, p} }{ \Delta_1  } \right) e^{ - \frac{\beta_{\rm int}}{2}( \Delta_{\ell, p} - \Delta_1 ) }\nonumber \\
G_{\infty} (\beta_{\rm int}) &= \sum_{ (\ell, p) \neq ( 1, 0 ) } \left( \frac{d_{\ell, p}}{d_1} \right) e^{ - \frac{\beta_{\rm int}}{2}( \Delta_{\ell, p} - \Delta_1 ) } , 
\end{align}
$d_1 \equiv d_{1,0}$, and $G_{\infty} ( \beta_{\rm int}) = \lim_{n \rightarrow \infty} ( F_n(\beta_{\rm int}) + D_n(\beta_{\rm int}) ) $.
Let us next consider the limit $\beta_{\rm int} \rightarrow \infty $. We can rewrite the correction term as
\begin{align}
G_{\infty} (\beta_{\rm int}) &= \sum_{ (\ell, p) \neq ( 1, 0 ) } \bar{b}_{\ell, p} e^{- ( \beta - \tau_0 ) \Delta_{p,0} - \frac{ \beta_{\rm int} }{ 2 } ( \Delta_{\ell, p} - \Delta_1 ) } \nonumber \\
&= \sum_{ (\ell, p) \neq ( 1, 0 ) } \bar{b}_{\ell, p} e^{- ( \beta - \tau_0 - \frac{\beta_{\rm int}}{2}) \Delta_{p,0} - \frac{ \beta_{\rm int} }{ 2 } \Delta_{\ell, 1 }} \nonumber \\
&\rightarrow 0 \qquad ( \beta_{\rm int} \rightarrow \infty ),
\end{align}
where
\begin{align}
\bar{b}_{\ell, p} = \frac{ \displaystyle \sum_q b_{\ell, p, q} e^{ - \tau_0 E_q } }{ \displaystyle \sum_q b_{1,0,q} e^{ - \tau_0 E_q } }.
\end{align}
Note that $\beta - \tau_0 - \frac{ \beta_{\rm int} }{2} \geq \frac{ \beta_{\rm int} }{ 2 } $ since $\beta \geq \tau_0 + \beta_{\rm int}$, $\Delta_{p,0} > 0$ $(p \neq 0)$, and $\Delta_{\ell, 1} > 0$ $(\ell > 1)$.  Similarly, $R_{\infty}(\beta_{\rm int}) \rightarrow 0 $ $(
\beta_{\rm int} \rightarrow \infty )$. Consequently,
\begin{align}
\lim_{\beta_{\rm int} \rightarrow \infty} \lim_{n \rightarrow \infty}
\hat{\Delta}_{(n, \beta_{\rm int})} = \Delta_1 .
\end{align} 
The convergence of $\beta_{\rm int} \rightarrow \infty$ limit is accelerated together with taking $\tau_0 \rightarrow \infty$ limit practically. Therefore we have analytically shown that the limits are interchangeable:
\begin{align}
\lim_{n \rightarrow \infty} \lim_{\beta_{\rm int}, \tau_0 \rightarrow \infty} \hat{\Delta}_{ (n,\beta_{\rm int}) } = \lim_{\beta_{\rm int}, \tau_0 \rightarrow \infty} \lim_{n \rightarrow \infty} \hat{\Delta}_{ (n,\beta_{\rm int}) } = \Delta_1 \nonumber .
\end{align}
This appealing property makes the estimation robust in practice and allows for flexibility in how the limit is taken.

As we have shown in the main text, we extrapolate the final gap
estimation for $n \rightarrow \infty$ first and for $\beta_{\rm int} \rightarrow \infty$ later. It is because that data for large
$\beta_{\rm int}$ will be noisy, and it is easier to take the limit
for $n \rightarrow \infty$ first. Nonetheless, we could invert the limits. Then let us consider the finite $n$ correction for making the extrapolation more reliable. Let $Q(n,z) =
\prod_{k=1}^n ( 1 + z^2 / k^2 )$. We have already used the limit
$Q(n,z) \rightarrow \sinh (\pi z) / \pi z$ $(n \rightarrow
\infty)$. The finite $n$ correction is expressed as
\begin{align}
& Q(n,z) \sim \frac{ \sinh ( \pi z ) }{ \pi z } \exp \left\{ - \frac{z^2}{n+1} - \frac{z^2}{2(n+1)(n+2) } \right\} \nonumber \\
& \mbox{\hspace{-0mm}} \sim \frac{ \sinh (\pi z) }{ \pi z } \left[ 1 - \frac{ z^2 }{n+1} - \frac{z^2}{2 (n+1) (n+2) } + \frac{ z^4}{2 (n+1)^2 } \right] \nonumber .
\end{align}
Here the asymptotic expansion of the Riemann zeta function was used;
\begin{align}
\sum_{k=1}^n \frac{ 1 }{ k^2 } = \frac{ \pi^2 }{ 6 } - \frac{ 1 }{ n + 1 } - \frac{ 1 }{ 2 (n+1) (n+2) } - \cdots .
\end{align}

As a result, the asymptotic systematic error of the gap estimator is rewritten for $\beta_{\rm int}, \tau_0 \gg 1 $ as
\begin{align}
\hat{\Delta}_{(n,\beta_{\rm int})} &\sim \Delta_1 + \sum_{ \ell > 1 } \left( \frac{ b_{\ell} }{ b_{1} } \right) \Delta_{\ell, 1} e^{- \frac{\beta_{\rm int} }{2} \Delta_{\ell, 1} } \label{eqn:gap-fit} \\
& \mbox{\hspace{-10mm}} \left[ 1 + \frac{ \bar{z}^2_{\ell} }{ n+1 } + \frac{ \bar{z}^2_{\ell} }{ 2 (n+1) } \left( \frac{ \bar{z}^2_{\ell} }{ n+1 } + \frac{1}{ n+2 } \right) + O \left( \frac{1}{n^3} \right) \right] \nonumber ,
\end{align}
where $b_{\ell} > 0$ $( \ell \geq 1 )$, ${\Delta}_{\ell, 1} \equiv
\Delta_{\ell} - \Delta_1 > 0$ and $\bar{z}^2_{\ell} = \left( \frac{
  \beta_{\rm int} }{ 2 \pi } \right)^2 ( \Delta^2_{\ell} - \Delta^2_1
) > 0$ $(\ell > 1)$. In practice, we can use a fitting function
$f(n,\beta_{\rm int}) = \Delta_1 + a \exp ( - b \, \beta_{\rm int} ) (
1 + c / n + d / n^2 )$ with positive real parameters $a, b, c$, and
real parameter $d$ to extrapolate the first gap, as demonstrated in
Fig.~\ref{fig:gap-n} and Fig.~\ref{fig:gap-beta} in the main
text. (Note that the parameter $d$ may be negative for small
$\beta_{\rm int}$.)

\end{document}